\documentclass{article}

\usepackage{iclr2027_conference,times}

\usepackage[utf8]{inputenc}
\usepackage[T1]{fontenc}
\usepackage{hyperref}
\usepackage{xcolor}

\usepackage{xurl}  
\usepackage{placeins}  

\usepackage{booktabs}
\usepackage{multirow}
\usepackage{amsmath}
\usepackage{amssymb}
\usepackage{graphicx}
\usepackage{algorithm}
\usepackage{algpseudocode}
\usepackage{siunitx}
\title{Optimizing the Optimizer: Language Models Discover
       Faster Molecular Relaxation Algorithms}

\iclrfinalcopy
\author{%
  Artem Tsypin\textsuperscript{1,2,\ensuremath{\dagger}} \quad
  Vladimir Deshchenya\textsuperscript{1,\ensuremath{\dagger}} \quad
  Kuzma Khrabrov\textsuperscript{1,2} \quad
  Denis Potapov\textsuperscript{1} \\
  \textbf{Maxim Radchenko}\textsuperscript{2} \quad
  \textbf{Artur Kadurin}\textsuperscript{1} \quad
  \textbf{Michael G. Medvedev}\textsuperscript{2,3} \\[0.3em]
  {\small \textsuperscript{1}AXXX \quad
  \textsuperscript{2}ITMO University} \\
  {\small \textsuperscript{3}N. D. Zelinsky Institute of Organic Chemistry of Russian Academy of Sciences} \\[0.2em]
  {\small \texttt{\{tsypin,deshchenya\}@axxx.tech}} \\[0.2em]
  {\small \textsuperscript{\ensuremath{\dagger}}Equal contribution.}
}

\begin{document}
\maketitle
\lhead{Preprint. Under review.}

\begin{abstract}
Geometry optimization is a major cost in many quantum-chemical workflows: each optimization step requires one force evaluation, and at the density-functional level that evaluation dominates the wall time.
Research in this area has produced a broad range of optimization methods, and we ask whether a language model can improve on the best of them through autoresearch.
An agent rewrites the optimizer itself to minimize force-call counts, restrained by two admission gates that reject premature stopping and improvements that do not generalize to unseen molecules.
Starting from Sella, the fastest open-source optimizer available, the search produces AutoSella, a family of two optimizers.
Both of them deliver consistent force-call reductions relative to Sella across held-out molecular benchmarks and potentials not used during the search.
Most notably, at the \texttt{r2SCAN-3c} DFT level, the best variant requires only $40.2$--$77.2\%$ of Sella's force calls while achieving the same energy reduction, even though agent used no DFT gradients.
\end{abstract}

\section{Introduction}

Molecular geometry optimization is one of the most frequently performed operations in computational chemistry, and its cost is often measured by the number of interatomic force evaluations.
On an inexpensive potential such as an empirical force field, that cost might be comparable to that of optimization algorithm itself.
However, at the density-functional theory (DFT) level, gradient calculations dominate the wall time of relaxation, so the number of steps required by an optimizer determines the price of optimization.
Reducing this number has been the subject of sustained work, from
quasi-Newton updates~\citep{nocedal1980, lbfgs1989}, rational-function
steps~\citep{rfo1985}, redundant internal coordinates~\citep{pulay1992} and model
Hessians~\citep{schlegel1984, fischer1992, lindh1995} to the coordinate
systems~\citep{bakerdlc1996, geometric2016}, graph-based
preconditioners~\citep{precon2016, preconff2018} and damped-dynamics
schemes~\citep{fire2020}.

Decades of manual algorithm design have shaped geometry-optimization software such as ASE~\citep{ase2017}, geomeTRIC~\citep{geometric2016, geometric2025}, pysisyphus~\citep{pysisyphus2021}, and Sella~\citep{sella2019, sella2021geodesic, sella2022}. Large language models (LLMs) now offer another way to search for improvements by proposing changes to optimizer code. Evolutionary program search with an LLM as
the mutation operator has recovered mathematical constructions and practical
kernels that improve on human designs~\citep{funsearch2024, alphaevolve2025, gigaevo2025}.
Agentic systems~\citep{saga2025, autoresearch2026, autoscientists2026} run semi- and fully automatic pipelines that evolve solutions against a machine-checkable objective such as wall time to reach the target validation loss~\citep{speedrun2024, speedrunbench2025}, compression quality under a hard artifact-size budget~\citep{paramgolf2026, paramgolfstudy2026}, or held-out accuracy on a suite of property prediction tasks~\citep{drugsage2026}.

\begin{figure}[t]
\centering
\includegraphics[width=\textwidth]{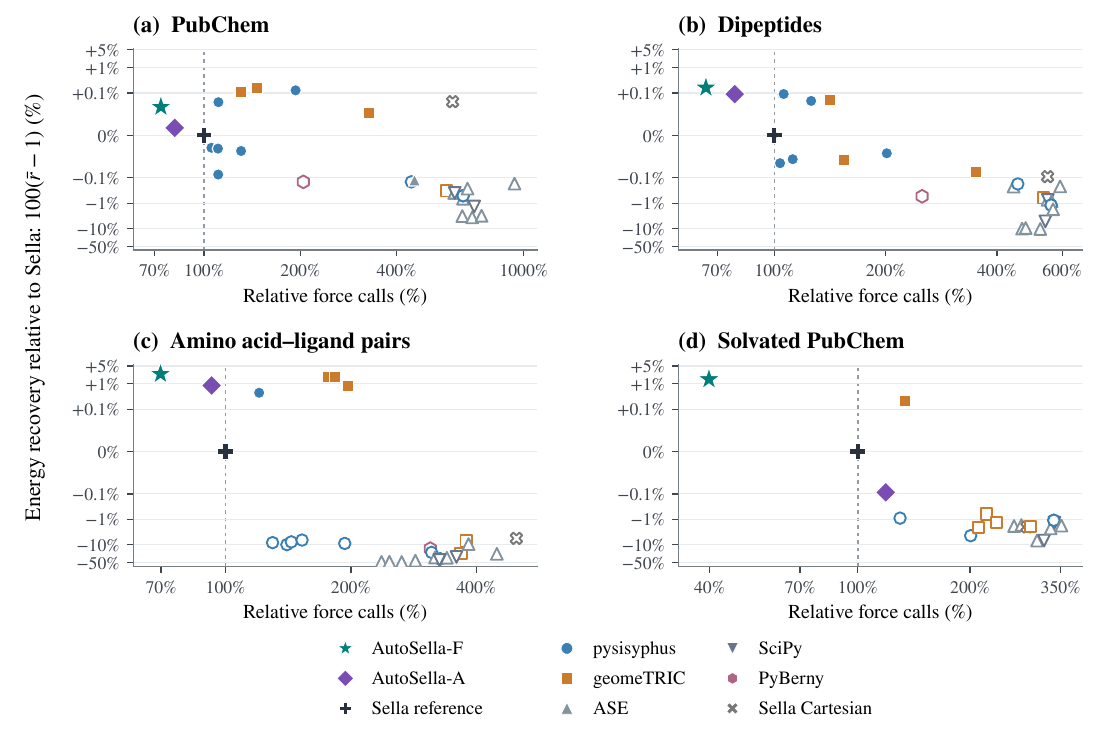}
\caption{\textbf{Count of force-calls against energy recovery on GFN2-xTB.}
One point per optimizer configuration: the two AutoSella variants and 27 configurations of five packages measured under an identical evaluation protocol.
Horizontal axis is the per-molecule relative cost $C_{\text{per-mol}}$ (Eq.~\eqref{eq:cost}).
The vertical axis shows \(100(\bar r-1)\), where \(\bar r\) (Eq.~\eqref{eq:energy}) is the mean recovered-energy ratio: zero denotes equal recovery, and positive values indicate greater energy lowering, on a symlog scale.
Open markers indicate failure or nonconvergence for more than 10\% of systems.
Full panel is Table~\ref{app:baselines}.
}
\label{fig:main}
\end{figure}

In this work, we apply the autoresearch paradigm to the design of molecular geometry optimizers.
Sella was developed at Sandia National Laboratories by Eric D. Hermes and colleagues and has been described in a series of research papers since 2019~\citep{sella2019,sella2021geodesic,sella2022}.
The project remained actively maintained in 2026, with version 2.5.0 released in June~\citep{sella2026v250}.
We show that Sella~2.5.0 required the fewest force calls among the 27 modern optimizer configurations we tested (Figure~\ref{fig:main}) and use its code as the starting point for our autoresearch runs.
An agent repeatedly proposes and applies edits aiming at producing candidates that use fewer force calls and reach
the same minima.

Our protocol addresses three constraints.
First, DFT relaxations are too costly to evaluate hundreds of optimizer candidates.
We therefore conduct the search on GFN2-xTB~\citep{gfn2xtb2019}, whose forces are accurate enough to guide the search and cheap enough for repeated evaluation (Section~\ref{sec:potential}).
We then demonstrate that the resulting improvements transfer to DFT (Section~\ref{sec:transfer}).
Second, candidate optimizers must generalize to unseen molecules.
A \emph{generalization gate} therefore evaluates promising edits on a validation split, requiring lower force-call cost without loss of relaxation depth (Section~\ref{sec:gates}).
Third, an optimizer can reduce force calls by simply stopping too early.
To prevent this, we added a \emph{validity gate} that rejects candidates with evaluation errors or mean relative energy recovery below Sella's (Section~\ref{sec:gates}).

Our contributions:
i) \emph{gated autoresearch loop} for improving molecular geometry optimizers (see Algorithm~\ref{alg:ar}, Section~\ref{sec:gates} and Appendix~\ref{app:autoresearch});
ii) \textbf{AutoSella}, a family of evolved optimizers --- \textbf{AutoSella-F} and \textbf{AutoSella-A} --- from two independent autoresearch runs with Claude Fable 5.1 (Max) and GPT-6 Astra (High), respectively\footnote{We also run an autoresearch experiment with Grok-4.6 (xhigh) and report results in Appendix~\ref{app:grok}.};
and iii) an extensive evaluation of those optimizers, testing generalization to unseen molecules, new chemical domain (Tables~\ref{tab:main},~\ref{tab:eval}), as well as to potentials the search never saw (Table~\ref{tab:eval}).

Our code is publicly available at \url{https://github.com/vdeshchenya/autosella}.

\section{Related work}
\label{app:related}

\paragraph{Language models that write algorithms.}
FunSearch~\citep{funsearch2024} evolves programs against a fixed scorer and AlphaEvolve~\citep{alphaevolve2025} extends the pattern to larger codebases and practical kernels.
A parallel line targets optimization heuristics: EoH~\citep{eoh2024} evolves a natural-language description of a heuristic together with its code, ReEvo~\citep{reevo2024} adds reflection over past attempts, LLaMEA~\citep{llamea2024} generates complete metaheuristics, and program search over an optimizer grammar predates all of them and produced Lion~\citep{lion2023}.
These approaches follow a pattern we also adopt in this work, a generator editing code against an evaluator it cannot modify, and differ in how the harness is set up.

\paragraph{Autonomous research loops.}
\texttt{autoresearch}~\citep{autoresearch2026} fixes a benchmark, restricts the agent to one file, time-boxes each experiment and keeps or discards on the measured result; the loop of Section~\ref{sec:autoresearch_loop} is an instance of it.
Other approaches automate even more of the research process: The AI Scientist~\citep{aiscientist2024, aiscientistv2} generates hypotheses, runs experiments and writes the manuscript, Agent Laboratory~\citep{agentlab2025} does so under human direction, AlphaLab~\citep{alphalab2026} and AutoSOTA~\citep{autosota2026} run multi-agent searches over optimization and model-design tasks, and similar method has been applied to reinforcement-learning algorithms~\citep{llmrlalgo2026}.
AutoScientists~\citep{autoscientists2026} replaces the single trajectory with a decentralized team that critiques proposals before spending compute and shares failed directions. Our loop is single-trajectory, making it significantly cheaper in terms of token usage.
DrugSAGE~\citep{drugsage2026} carries the same accumulated experience multi-agent research into chemistry, reusing it across property-prediction tasks.
SAGA~\citep{saga2025} argues that the objectives given to discovery agents are imperfect proxies and evolves the objective in an outer loop.
We instead engineer a set of gates alongside the main objective to prevent reward hacking and overfitting.

\paragraph{Relaxation with MLFFs.}
The advent of universal MLFFs has made large-scale molecular geometry optimization possible at near-DFT accuracy.
A set of benchmarks has emerged to measure the quality of relaxation with MLFFs: GOLF~\citep{tsypin2024gradual}, $\nabla^2$DFT~\citep{khrabrov2024nabladft} and PubChemQCR~\citep{pubchemqcr2025} curate DFT optimization trajectories of organic molecules and evaluate potentials by running BFGS relaxations under them.
Matbench Discovery~\citep{matbench2025} and OC20~\citep{oc20} rank potentials based on the structures their relaxations reach.
Another benchmark~\citep{rowansci2025} compares different optimizers on a fixed set of molecules with a given potential.
We develop a new optimizer whose reduction in force calls persists across different potentials (Section~\ref{sec:transfer}), complementing the benchmarks described above.

\section{Method}
\label{sec:method}

\subsection{Task and metrics}
\label{sec:task}
An optimizer is handed a molecule and a force evaluator and must reach the
fixed convergence criteria in as few force calls as possible. We write $n_i$ for the force calls it uses on molecule $i$, and $n_i^{\mathrm{ref}}$ for the calls the reference optimizer uses on the same molecule. Cost is reported as two aggregates of that ratio over a benchmark of $N$ molecules,
\begin{equation}
  C_{\mathrm{pooled}} \;=\; \frac{\sum_i n_i}{\sum_i n_i^{\mathrm{ref}}},
  \qquad\qquad
  C_{\text{per-mol}} \;=\; \frac{1}{N}\sum_i \frac{n_i}{n_i^{\mathrm{ref}}},
  \label{eq:cost}
\end{equation}
which differ in the weight a long trajectory carries: $C_{\mathrm{pooled}}$ is
dominated by the molecules the reference spends most force calls on, while
$C_{\text{per-mol}}$ treats every molecule evenly.
Relaxation depth is scored by the relative energy recovery
\begin{equation}
  r_i \;=\; \frac{E^{\mathrm{start}}_i - E^{\mathrm{final}}_i}
                 {E^{\mathrm{start}}_i - E^{\mathrm{final,ref}}_i},
  \qquad\qquad
  \bar r \;=\; \frac{1}{N}\sum_i r_i,
  \label{eq:energy}
\end{equation}
with $r_i$ equal to $1$ when the optimizer descends exactly as far as the reference, less than $1$ when it stops short, and greater than $1$ when it reaches a deeper minimum.

\subsection{Molecules}
\label{sec:molecules}
In this work, we focus on geometry optimization for drug design, considering both drug-like molecules and their interactions with surrounding biomolecules and solvent.
We use SPICE 2.0.1~\citep{spice2023} as the main source of structures for training, validation, and evaluation.
For the training and validation sets, we select drug-like molecules from its PubChem subset and dimers from the DES370K subset~\citep{des370k2021} to improve optimization cost on non-covalent interactions.
We use diversity-based selection procedures to draw 225 PubChem molecules and 219 DES370K dimers for training, and 250 PubChem molecules and 215 DES370K dimers for validation. We supplement the training set with 25 registered drugs from a published optimizer benchmark~\citep{rowansci2025}, yielding 469 training and 465 validation structures.

Evaluation uses four held-out datasets.
Two test sets focus on single-molecule systems: 500 out-of-distribution molecules selected from the PubChem subset and 673 structures from Dipeptides.
The former tests generalization to drug-like molecules outside the training and validation sets, while the latter probes transfer to another chemical domain.
A third set contains 475 systems from Amino Acid Ligand Pairs and tests optimization of non-covalent contacts, extending beyond the small DES370K dimers used during training.
Finally, 100 systems from Solvated PubChem test complex, multi-fragment optimization, where relaxation involves coupled rearrangements of a drug-like solute and surrounding water molecules.
The detailed data preparation pipeline is in Appendix~\ref{app:data}.

\subsection{Potential and convergence criteria}
\label{sec:potential}
\label{sec:reference}
The software settings and evaluation infrastructure are described in Appendix~\ref{app:harness}.
Evolution is performed at the GFN2-xTB level~\citep{gfn2xtb2019, xtbreview2021}, which balances forces' accuracy and relatively low cost.
A full evaluation of a single optimizer candidate on the 469-structure training set takes approximately 60--90\,s of wall time on average, depending on the autoresearch experiment.\footnote{Timed on four Intel Xeon Gold 6348, running 48 distributed workers each.}
Convergence is the five-criterion Gaussian-style test used by geomeTRIC~\citep{geometric2016}.
Writing $\Delta E$ for the energy change between successive force evaluations,
$g$ for the gradient and $\Delta x$ for the displacement, an optimization converges only
when all five conditions hold at once:
\begin{equation}
\begin{aligned}
  |\Delta E| &< 10^{-6}\,E_\mathrm{h}, &
  \max|g| &< 4.5\times10^{-4}\,E_\mathrm{h}\,a_0^{-1}, &
  \operatorname{rms} g &< 3\times10^{-4}\,E_\mathrm{h}\,a_0^{-1}, \\[2pt]
  \max|\Delta x| &< 1.8\times10^{-3}\,\text{\AA}, &
  \operatorname{rms} \Delta x &< 1.2\times10^{-3}\,\text{\AA}. & &
\end{aligned}
\label{eq:conv}
\end{equation}

Maximum and RMS gradient and displacement values are computed from the Euclidean norms of the corresponding per-atom vectors.
Displacements are measured between consecutive evaluated geometries after optimal rigid-body alignment to remove overall translation and rotation.

Training and validation optimizations are capped at 200 force calls.
This limit was selected using the training and validation sets and is deliberately loose: the Sella reference converges on every structure in these splits, using 29 force calls on average and at most 155.
During training and validation, reaching a force-call or time limit without convergence does not by itself invalidate the submitted algorithm (see Section~\ref{sec:gates}).
Held-out optimizations use the same 200-call cap, increased to 500 for Solvated PubChem because of the greater complexity of these systems.
Sella converges on all 1,648 structures in the PubChem test, Dipeptides, and amino acid–ligand datasets, using 38 force calls on average and at most 200 with GFN2-xTB.
On Solvated PubChem, Sella converges on 98 of 100 systems, using 209 force calls on average and at most 366.
Of the remaining two systems, one reaches the 500-call cap and the other fails during Sella’s geometry update.

\subsection{The autoresearch loop} \label{sec:autoresearch_loop}
\label{sec:loop}
We chose Sella~2.5.0~\cite{sella2022, sella2026v250} as the starting point for the evolution.
We tested the majority of open-source optimizers that support custom convergence criteria defined in Eq.~\eqref{eq:conv}, and found Sella to use the fewest force calls on all our benchmarks (Figure~\ref{fig:main}, Appendix~\ref{sec:classical-panel}).\footnote{27 configurations of optimizers from six packages, described in Appendix~\ref{sec:optimizer-list}}
To enable integration of Sella into the autoresearch loop, we consolidated its codebase into a single file (see Appendix~\ref{app:seed}).

The evolution follows the autoresearch pattern~\citep{autoresearch2026}: a fixed benchmark, a single editable file and a keep-or-discard decision (see Section~\ref{sec:gates}) on the measured result.
An agent repeatedly edits \texttt{algo.py}, the single file it may modify, commits the change, and has the updated version scored on the training split by an evaluator.
The quantity being minimized is the force-call cost defined in Eq.~\eqref{eq:cost}.
We write $a^\star$ for the \emph{champion}, the best optimizer so far, and $a^t$ for the candidate obtained by editing it at step $t$.
The evaluator returns the candidate's relative force calls $C^t_{\text{per-mol}}$ and energy recovery $\bar r^t$ on the training split and $C^{t, V}_{\text{per-mol}}$, $\bar r^{t, V}$ on the validation split.
After every evaluation, the agent reads a \emph{per-molecule} table which contains relative force calls, relative energy, final energy difference and convergence status for each molecule.
This supports formulation of chemically grounded hypotheses, since a mechanism can be targeted at specific molecules.
The Algorithm~\ref{alg:ar} and Appendix~\ref{app:autoresearch} describe the loop in detail.

\begin{algorithm}[!t]
\small
\caption{The autoresearch loop, in the notation of Section~\ref{sec:gates}.
$\mathcal{E}(a)$ and $\mathcal{E}^V(a)$ evaluate program $a$ on the training
and validation splits, returning its per-molecule cost $C_{\text{per-mol}}$
(Eq.~\eqref{eq:cost}) and mean energy recovery $\bar r$
(Eq.~\eqref{eq:energy}).
The current champion is $a^{\star}$.
A discarded edit leaves no trace in the program, only
in the record.
Every branch below is computed from $\mathcal{E}$; the identity-gating rule on line~\ref{ln:propose} is not, and is applied by the agent to its own proposal.}
\label{alg:ar}
\begin{algorithmic}[1]
\Require training and validation splits, evaluator $\mathcal{E}$,
         significance threshold $\delta = 10^{-4}$
\State $a^\star \gets$ vendored Sella~2.5.0 \Comment{Appendix~\ref{app:seed}}
\State $(C^\star_{\text{per-mol}}, \cdot) \gets \mathcal{E}(a^\star)$; \quad
       $(C^{\star,V}_{\text{per-mol}}, \cdot) \gets \mathcal{E}^V(a^\star)$ \Comment{The cost is 1.0 for the baseline}
\State $t \gets 0$
\While{\textbf{true}}
  \State $t \gets t+1$
  \State read the cycle record, the narrative log, and the mechanisms already
         rejected
  \State $a^t \gets$ one coherent edit to $a^\star$ \label{ln:propose}
  \State $(C^t_{\text{per-mol}}, \bar r^t) \gets \mathcal{E}(a^t)$
  \State append the result to the cycle record and a narrative section to the
         log \Comment{crashes and rejects included}
  \If{$\bar r^t < 1$ \textbf{or} any molecule has an evaluation error}
    \State discard $a^t$ \Comment{validity gate: faster but shallower is not
           faster}
  \ElsIf{$C^t_{\text{per-mol}} \ge C^\star_{\text{per-mol}} - \delta$}
    \State discard $a^t$ \Comment{not a significant improvement on training}
  \Else
    \State $(C^{t,V}_{\text{per-mol}}, \bar r^{t,V}) \gets \mathcal{E}^V(a^t)$
           \Comment{generalization gate}
    \If{no evaluation errors \textbf{and} $\bar r^{t,V} \ge 1$ \textbf{and}
        $C^{t,V}_{\text{per-mol}} < C^{\star,V}_{\text{per-mol}} - \delta$}
      \State $a^\star \gets a^t$; \quad
             $C^\star_{\text{per-mol}} \gets C^t_{\text{per-mol}}$; \quad
             $C^{\star,V}_{\text{per-mol}} \gets C^{t,V}_{\text{per-mol}}$
             \Comment{accept as the new champion}
    \Else
      \State discard $a^t$ and record the mechanism as non-generalizing
    \EndIf
  \EndIf
\EndWhile
\end{algorithmic}
\end{algorithm}

We ran the loop independently for 66 active research hours with Claude Code driving Claude Fable~5.1 (max) and Codex driving GPT-6 Astra (high), producing \textbf{AutoSella-F} and \textbf{AutoSella-A}, respectively.
The autoresearch harness, including the isolated container and remote evaluation setup, is described in Appendix~\ref{app:harness}.

\subsection{Candidate acceptance criteria}
\label{sec:gates}

A candidate $a^t$ is evaluated on training first.
It must have $\bar r^t \ge 1$, no evaluation errors such as optimizer exceptions, non-finite numerical values, malformed output, and a force-call cost below $C^\star_{\text{per-mol}}-\delta$, where $\delta=10^{-4}$ keeps numerical noise from counting as progress.
Note that on training and validation, non-converged optimizations stopped by a force-call remain in the averages.
We call these conditions the \emph{validity gate}.
If they are met, $a^t$ is evaluated on the validation split and replaces the champion only if the same requirements hold there: no evaluation errors, $\bar r^{t,V}\ge1$, and $C^{t,V}_{\text{per-mol}}<C^{\star,V}_{\text{per-mol}}-\delta$ -- the \emph{generalization gate}.
If either gate fails, the edit is discarded and the agent edits $a^\star$ again.
The condition $\bar r^t \ge 1$ stops the agent from hacking the convergence criteria and saving force calls by under-relaxing.
The generalization gate then ensures that accepted improvement holds on unseen molecules, preventing the overfitting to the training set.
Neither gate inspects the geometry itself, so we additionally compare the optimized conformations with the ones from Sella, by heavy-atom RMSD and by vibrational analysis, in Appendix~\ref{app:conf}.
Figure~\ref{fig:progress} illustrates the autoresearch progress and how the gates reject candidates.

\section{Results}
\label{sec:results}
We test the AutoSella optimizers along two generalization axes.
\emph{Molecules:} four held-out datasets comprising 500 PubChem molecules selected to be dissimilar from the training and validation sets, 673 dipeptides, 475 amino acid--ligand pairs, and 100 Solvated PubChem systems.
Amino acid--ligand pairs test generalization of optimization to systems with non-covalent interactions beyond the small dimers used in training, while Solvated PubChem tests optimization of complex, multi-fragment systems involving a drug-like solute and surrounding water molecules.
\emph{Potential:} GFN2-xTB, which the autoresearch was performed with, and two potentials it never encountered: the GFN-FF force field~\citep{spicher2020gfnff} and the \texttt{r2SCAN-3c} DFT functional.
We also analyze the autoresearch progress for both AutoSella variants in Section~\ref{sec:ar-progress}.

\begin{table}[!t]
\centering
\caption{\textbf{Force-call cost and energy fidelity on GFN2-xTB, the potential the optimizers were evolved on.} $C_{\mathrm{pooled}}$, $C_{\text{per-mol}}$ and $\bar r$ are defined in Eqs.~\eqref{eq:cost} and~\eqref{eq:energy}. \emph{Err. b/a}\ counts optimization errors and molecules not converged in 200 steps (500 steps for Solvated Pubchem) for \emph{baseline}/\emph{algorithm}; means exclude them.}
\label{tab:main}
\small
\setlength{\tabcolsep}{3pt}
\renewcommand{\arraystretch}{0.98}
\setlength{\aboverulesep}{1.5pt}
\setlength{\belowrulesep}{1.5pt}
\begin{tabular*}{0.94\textwidth}{@{\extracolsep{\fill}}l r l c S[table-format=3.2] S[table-format=3.2] S[table-format=1.6, group-digits=none]@{}}
\toprule
\multicolumn{4}{@{}l}{} & \multicolumn{2}{c}{Relative force calls (\%, Eq.~\eqref{eq:cost})\,$\downarrow$} & \\
\cmidrule(lr){5-6}
Benchmark & {$N$} & Method & {Err.\ b/a\,$\downarrow$} & {$C_{\mathrm{pooled}}$} & {$C_{\text{per-mol}}$} & {$\bar r$\,$\uparrow$} \\
\midrule
\multirow{2}{*}{Test} & \multirow{2}{*}{500} & AutoSella-F & 0/0 & \bfseries 72.05 & \bfseries 73.25 & \bfseries 1.000284 \\
 &  & AutoSella-A & 0/1 & 80.34 & 80.94 & 1.000044 \\
\cmidrule(lr){1-7}
\multirow{2}{*}{Dipeptides} & \multirow{2}{*}{673} & AutoSella-F & 0/0 & \bfseries 64.06 & \bfseries 65.28 & \bfseries 1.001613 \\
 &  & AutoSella-A & 0/0 & 77.01 & 78.14 & 1.000908 \\
\cmidrule(lr){1-7}
\multirow{2}{*}{AA--ligand pairs} & \multirow{2}{*}{475} & AutoSella-F & 0/0 & \bfseries 67.21 & \bfseries 69.79 & \bfseries 1.024184 \\
 &  & AutoSella-A & 0/1 & 90.41 & 92.57 & 1.008460 \\
\cmidrule(lr){1-7}
\multirow{2}{*}{Solvated PubChem} & \multirow{2}{*}{100} & AutoSella-F & 2/0 & \bfseries 38.95 & \bfseries 39.86 & \bfseries 1.014870 \\
 &  & AutoSella-A & 2/0 & 113.01 & 118.77 & 0.999125 \\
\bottomrule
\end{tabular*}
\end{table}

\subsection{Cost on GFN2-xTB}
\label{sec:cost-xtb}
Table~\ref{tab:main} compares both AutoSella variants with Sella across the four held-out datasets.
AutoSella-F is cheapest in every block, at $39.86$--$73.25\%$ of the baseline's force calls per molecule.
Its mean energy recovery is above the baseline on all four datasets ($\bar r>1$).
Appendix~\ref{app:dE} analyses that difference in detail: the median $\Delta E = E^{\mathrm{AutoSella-F}}_{\mathrm{final}} - E^{\mathrm{Sella}}_{\mathrm{final}}$ ranges from $-8\times10^{-6}$ to $-5\times10^{-6}$\,kcal\,mol$^{-1}$ on Test, Dipeptides, and AA--ligand pairs, and is $-1.30$\,kcal\,mol$^{-1}$ on Solvated PubChem, so AutoSella-F reaches lower final energies than the reference on most molecules.
The saving therefore holds across unseen molecules and new chemical domains, on the potential the search itself was run on.
AutoSella-A uses $78.14$--$92.57\%$ of the baseline's force calls per molecule on the first three datasets, but regresses on Solvated PubChem, requiring $118.77\%$ while recovering less energy ($\bar r=0.999125$).

\begin{table}[!t]
\centering
\caption{\textbf{Transfer to two potentials never seen during evolution.} GFN-FF, and \texttt{r2SCAN-3c}. All columns except for ``Potential'' are identical to Table~\ref{tab:main}.
}
\label{tab:eval}
\small
\setlength{\tabcolsep}{2pt}
\renewcommand{\arraystretch}{0.98}
\setlength{\aboverulesep}{1.5pt}
\setlength{\belowrulesep}{1.5pt}
\begin{tabular*}{\textwidth}{@{\extracolsep{\fill}}l l r l c S[table-format=3.2] S[table-format=3.2] S[table-format=1.6, group-digits=none]@{}}
\toprule
\multicolumn{5}{@{}l}{} & \multicolumn{2}{c}{Relative force calls (\%, Eq.~\eqref{eq:cost})\,$\downarrow$} & \\
\cmidrule(lr){6-7}
Potential & Benchmark & {$N$} & Method & {Err.\ b/a\,$\downarrow$} & {$C_{\mathrm{pooled}}$} & {$C_{\text{per-mol}}$} & {$\bar r$\,$\uparrow$} \\
\midrule
\multirow{8}{*}{GFN-FF} & \multirow{2}{*}{Test} & \multirow{2}{*}{500} & AutoSella-F & 0/1 & \bfseries 74.31 & \bfseries 74.14 & \bfseries 1.002850 \\
 &  &  & AutoSella-A & 0/1 & 78.52 & 78.90 & 1.002606 \\
\cmidrule(lr){2-8}
 & \multirow{2}{*}{Dipeptides} & \multirow{2}{*}{673} & AutoSella-F & 0/0 & \bfseries 67.91 & \bfseries 68.91 & 0.999653 \\
 &  &  & AutoSella-A & 0/0 & 77.06 & 77.80 & \bfseries 1.000674 \\
\cmidrule(lr){2-8}
 & \multirow{2}{*}{AA--ligand pairs} & \multirow{2}{*}{475} & AutoSella-F & 2/1 & \bfseries 70.41 & \bfseries 72.57 & \bfseries 1.017843 \\
 &  &  & AutoSella-A & 2/1 & 86.63 & 88.68 & 0.997444 \\
\cmidrule(lr){2-8}
 & \multirow{2}{*}{Solvated PubChem} & \multirow{2}{*}{100} & AutoSella-F & 1/0 & \bfseries 33.34 & \bfseries 34.08 & \bfseries 1.012442 \\
 &  &  & AutoSella-A & 1/1 & 109.63 & 114.59 & 0.999054 \\
\midrule
\multirow{8}{*}{\texttt{r2SCAN-3c}} & \multirow{2}{*}{Test} & \multirow{2}{*}{500} & AutoSella-F & 0/0 & \bfseries 75.98 & \bfseries 77.20 & 1.000259 \\
 &  &  & AutoSella-A & 0/0 & 82.12 & 82.87 & \bfseries 1.000389 \\
\cmidrule(lr){2-8}
 & \multirow{2}{*}{Dipeptides} & \multirow{2}{*}{673} & AutoSella-F & 0/0 & \bfseries 72.27 & \bfseries 73.14 & 1.000111 \\
 &  &  & AutoSella-A & 0/0 & 77.87 & 78.43 & \bfseries 1.000485 \\
\cmidrule(lr){2-8}
 & \multirow{2}{*}{AA--ligand pairs} & \multirow{2}{*}{475} & AutoSella-F & 0/0 & \bfseries 65.50 & \bfseries 68.61 & \bfseries 1.028167 \\
 &  &  & AutoSella-A & 0/0 & 84.32 & 86.71 & 1.012738 \\
\cmidrule(lr){2-8}
 & \multirow{2}{*}{Solvated PubChem} & \multirow{2}{*}{100} & AutoSella-F & 0/0 & \bfseries 38.69 & \bfseries 40.24 & \bfseries 1.012674 \\
 &  &  & AutoSella-A & 0/0 & 105.75 & 109.06 & 0.993028 \\
\bottomrule
\end{tabular*}
\end{table}

\subsection{Transfer to unseen potentials}
\label{sec:transfer}
The autoresearch scored every candidate on GFN2-xTB alone, so the question is whether the mechanisms behind the improvement are general or fitted to that potential.
Table~\ref{tab:eval} repeats the measurement on the classical force field GFN-FF~\citep{spicher2020gfnff}, and on DFT at the \texttt{r2SCAN-3c} level~\citep{r2scan3c2021} through
ORCA~\citep{orca2022}.
We show that the force call saving carry over to them.
AutoSella-F saves more force calls than AutoSella-A in every available comparison.
The largest saving under DFT occurs on Solvated PubChem, where AutoSella-F requires only $40.24\%$ of Sella’s force calls while achieving $1.27\%$ greater mean energy recovery. Its mean energy recovery exceeds the baseline on all four datasets.
Consistent with the limited generalization to complex multi-fragment systems observed in Section~\ref{sec:cost-xtb}, AutoSella-A uses $109.06\%$ of Sella's force calls per molecule and recovers less energy ($\bar r=0.993028$) on Solvated PubChem systems under DFT.
Under DFT the force call dominates the calculation, so the reduction in force calls lowers the cost of the job proportionally, making the AutoSella optimizers practically useful.
Remarkably, this was obtained without the autoresearch ever evaluating a DFT gradient.

\begin{figure}[!t]
\centering
\includegraphics[width=\textwidth]{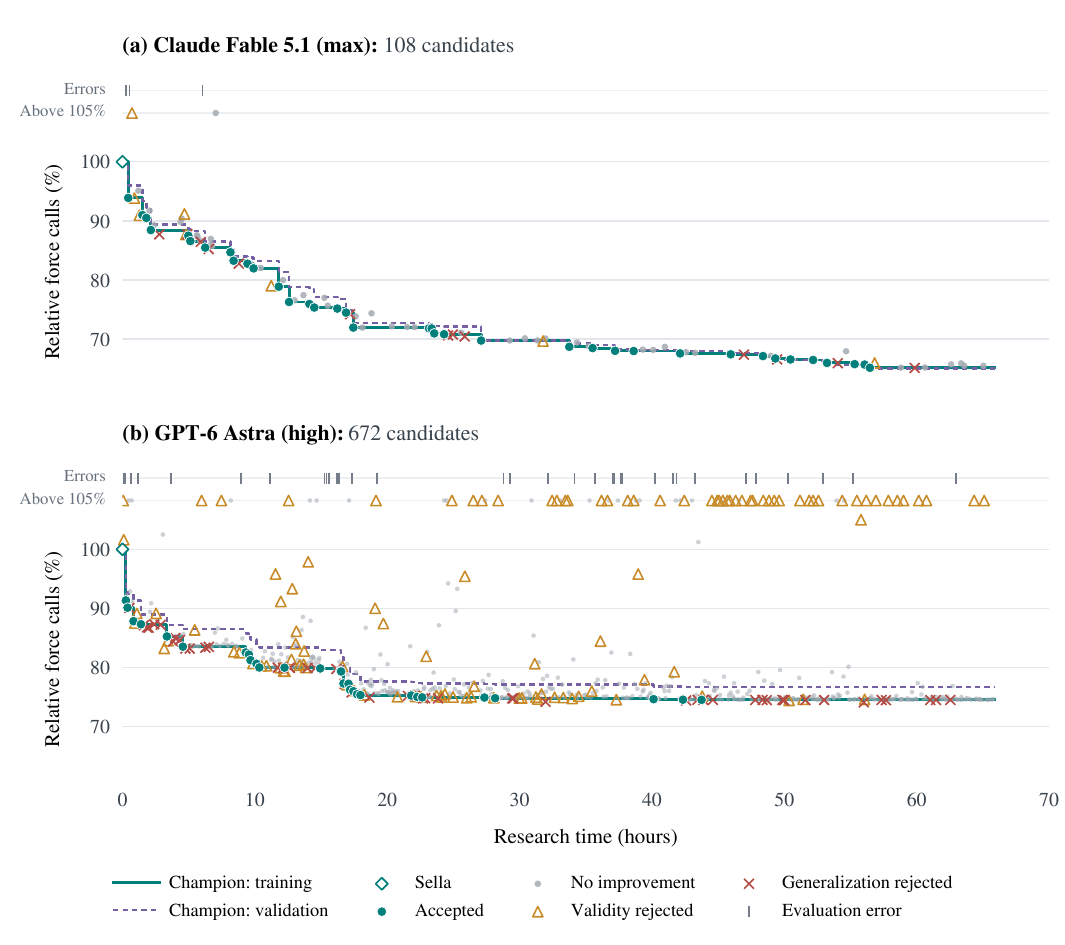}
\caption{
\textbf{Autoresearch progress of Claude Fable~5.1 (max) and GPT-6 Astra (high) over 66 active research hours.}
Solid and dashed lines show the current champion's training and validation force-call costs, relative to the initial Sella program.
Markers indicate candidate outcomes decided by admission gates of Algorithm~\ref{alg:ar}.
}
\label{fig:progress}
\end{figure}

\subsection{Autoresearch progress}
\label{sec:ar-progress}
In this section, we analyze the autoresearch progress for Claude Fable 5.1 (max) and GPT-6 Astra (high).
As different agents tend to produce candidates at a different pace, we use the active research time (in hours) to measure progress.
Figure~\ref{fig:progress} compares two LLMs over 66 active research hours.
This budget includes development and analysis in an isolated container as well as remote evaluation (see Section~\ref{app:harness}).
Within this window, Fable commits 108 evaluated candidates and accepts 37 improvements, whereas Astra commits 672 candidates and accepts 29.
Fable has fewer evaluation errors (3, or 2.8\%, versus 35, or 5.2\%) and fewer candidates rejected for insufficient mean energy recovery (8, or 7.4\%, versus 115, or 17.1\%).
The final algorithms, AutoSella-F and AutoSella-A, have mean per-molecule force-call ratios of 65.13\%/74.51\% on training and 64.99\%/76.66\% on validation, relative to the original Sella.

The spacing of commits demonstrates different research rhythms:
the mean gap between consecutive candidate commits is $36.32\pm32.08$ minutes for Fable and $5.88\pm3.66$ minutes for Astra (mean $\pm$ sample standard deviation; 107 and 671 intervals).
Fable's logs document extended local analysis inside its isolated container.
During the 160-minute interval between cycles 87 and 88, the agent used hundreds of shell commands and file reads to examine molecular structures saved from earlier calculations and inspect the optimizer code.
This analysis led to a targeted change in how optimization is initialized.
Thus the smaller number of candidates does not imply less research.

An analysis of Fable’s and Astra’s research histories reveals a similar pattern: both models propose new ideas, test their effects, diagnose failures, and successfully repair partially broken mechanisms.
Fable's research history suggests that its larger gains came from combining corrections to the physical model with more effective reuse of previous force evaluations.
For example, it added missing curvature from non-covalent contacts and updated this model as the geometry changed (cycles~34--35).
It then found that stored gradient differences could partly undo these corrections and adjusted their reuse to the current physical model (cycles~53--55).
Separate analyses produced inexpensive preparation of molecular conformations and fragment positions before the first xTB call, shortening the subsequent optimization.
Astra's later accepted gains were nevertheless small: many subsequent variants increased force-call cost or passed training but failed the validation cost or energy tests.

Appendix~\ref{app:interpretation} describes the optimizer improvements and evaluates their effects by removing groups of related changes from the final algorithms.
We highlight the three improvement groups (blocks) per model whose removal produces the largest increases in validation force-call cost (Figure~\ref{fig:family-ablations}).
For both optimizers, the two largest increases come from separately removing the model Hessian and coordinates block and the fragment handling block: approximately $18.7$ and $11.3$ percentage points for AutoSella-F, and $8.8$ and $5.7$ points for AutoSella-A, respectively.
The third-largest increase comes from removing Hessian updates in AutoSella-F ($10.2$ points) and the Gaussian process in AutoSella-A ($4.7$ points).

\section{Conclusion}
We applied the autoresearch paradigm to molecular geometry optimization in order to produce algorithms that use fewer force calls while preserving relaxation quality.
Starting from Sella, the fastest open-source optimizer in our benchmark, two independent autoresearch runs on GFN2-xTB driven by Claude Fable~5.1 (max) and GPT-6 Astra (high) produced AutoSella-F and AutoSella-A, respectively.
AutoSella-F achieved larger and more consistent savings, which transferred to GFN-FF and \texttt{r2SCAN-3c} DFT despite neither potential being used during the search.
Across four out-of-distribution datasets at the \texttt{r2SCAN-3c} level, it required approximately $23$--$60\%$ fewer force calls per molecule than Sella while achieving greater mean energy recovery.
Ablations highlight three key contributors to AutoSella-F's savings: improved curvature models and coordinates, explicit handling of disconnected fragments, and Hessian updates that reuse previous gradient information while accounting for changes in geometry.
These results suggest that inexpensive surrogate potentials can enable automated algorithm development when direct search at the target level is too costly.

\section*{AI use statement}
We used large language models and AI agents to build the autoresearch (AR) harness, assist with implementing and debugging code and prompts, and run autoresearch experiments and evaluations.
Within the AR harness, AI agents actively generated research ideas and developed the improved molecular optimization algorithm presented in this paper.
We also used LLMs to retrieve and identify related work, draft sections of the manuscript, and revise and polish the writing.
The authors take full responsibility for the final content of this paper, including its claims, results, references, and AI-assisted text, code, and other research artifacts.

%
%
\bibliographystyle{iclr2027_conference}
\bibliography{optimizers}

\newpage
\appendix
\FloatBarrier
\section{Data preparation}
\label{app:data}
We prepare the training, validation, and held-out evaluation inputs from SPICE~2.0.1~\citep{spice2023} and the fixed 25-structure RowanSci benchmark~\citep{rowansci2025}. The RowanSci coordinates are additionally perturbed by noise with a mean per-atom displacement of $0.02\,\text{\AA}$.

Before accepting a candidate, we relax it with Sella~2.5.0/GFN2-xTB to reject unstable or nearly stationary starting structures. For each SPICE system, we try up to ten random conformers and retain the first that passes the preparation checks. Acceptance requires an energy decrease of at least $0.005\,E_\mathrm{h}$, final connectivity consistent with the source mapped SMILES up to covalent-radius boundary ambiguities, and a final maximum interatomic distance no more than twice that of the input. Convergence within the force-call budget is not required. The budget is 200 calls, except for Solvated PubChem, where it is 500. We save the starting geometries, not the relaxed ones.

For the training and validation PubChem split, we deduplicate SPICE PubChem groups by canonical SMILES and compute 1024-bit Morgan fingerprints of radius~2 (ECFP4). Starting with the lowest-ID group that passes preparation, we greedily choose the remaining group with the smallest maximum Tanimoto similarity to previously accepted molecules. This promotes structural diversity among the selected molecules. The first 225 accepted groups form the PubChem training subset and the next 250 form validation. For DES370K~\citep{des370k2021}, we group dimers by their monomer chemical-class pair and seek one accepted training dimer per represented pair. We then seek a distinct validation dimer for each successful pair, preferring candidates without monomers already represented in the complete training selection. This yields 219 training and 215 validation dimers. The 25 RowanSci structures are assigned to training. The resulting sets contain 469 training and 465 validation structures.

The held-out PubChem Test set contains 500 additional accepted SPICE molecules. It uses the same 1024-bit fingerprint and greedy selection, initialized with the training and validation PubChem molecules, selected DES monomers, and RowanSci molecules. The Dipeptides set comprises the 673 accepted systems among the 676 fixed SPICE dipeptide identities. For Amino Acid Ligand Pairs, a stratified selection yields 25 accepted systems for each of 19 amino acids, with no ligand identifier reused across amino acids, for 475 systems total. Solvated PubChem contains 100 accepted SPICE systems, each supplied as one solute with 20 waters. We select them by the solute's SMILES and fingerprint, excluding solutes matching the main-set reference molecules and favoring those dissimilar to the references and previously accepted solvated solutes.

\section{Reference optimizer}
\label{app:seed}
Evolution starts from a single-file version of Sella, obtained by the coding agent and thoroughly tested for reproducing orginal sella's results. 
Compacting sella into a single file serves two purposes: the implementation can be cloudpickled by value to the evaluation workers, and its source is pinned independently of subsequent upstream releases.
The single file was produced from the official \texttt{sella}~2.5.0 package in four steps.

\textbf{Trace.}
The executed path was measured: A \texttt{sys.settrace} run recorded which functions \texttt{Sella(atoms, internal=True, order=0)} with \texttt{irun(fmax=0)} actually calls. Of the package's 9136 lines, that path touches 1418 lines and 131 functions across eight modules. Two consequences follow at once: i) \texttt{eigensolvers.py} never runs, as it is saddle-point machinery, and ii) the compiled \texttt{modified\_gram\_schmidt} has no call site, so the minimization path carries no Cython dependency.

\textbf{Concatenate.}
Those eight modules were joined in dependency order with intra-package imports stripped.
The result is the source file, which is numerically identical to the package by construction.

\textbf{Prune.}
Two passes followed: a static reachability pass from \texttt{Sella} and the entry point, then a fixpoint pass that drops untraced definitions and restores any the surviving code still names, repeated until stable. 
The restore step is necessary.
Without it the file fails at import, because definitions are reached through attribute assignment and through \texttt{jax} transformations applied in class bodies, neither of which a plain search over names finds.

\textbf{Collapse the device layer.}
The evaluation workers have no \texttt{torch}, so the GPU predicate is permanently false and every device branch is unreachable for any input.

The seed contains 5975 source lines, including comments, blank lines, and the evaluation wrapper, and is identical at the start of the Fable and Astra searches.

Sella's \texttt{allow\_fragments} argument controls the coordinate representation: \texttt{True} gives disconnected fragments explicit translation and rotation coordinates, whereas \texttt{False} joins them through artificial coordinate connections without changing the potential.
Testing both settings showed that \texttt{False} achieved greater energy reduction but required more force calls.
We therefore started the evolution with this more costly \texttt{allow\_fragments=False} variant, seeking efficiency gains while preserving its energy recovery.

The initial Sella program with \texttt{allow\_fragments=False} is the reference for the Fable/Astra and Grok progress plots (Figures~\ref{fig:progress} and~\ref{fig:progress-grok}) and the ablation study (Figure~\ref{fig:family-ablations}).
The evaluation comparisons use Sella with \texttt{allow\_fragments=True}: Figure~\ref{fig:main}, and Tables~\ref{tab:main}, \ref{tab:eval}, \ref{tab:dE}, \ref{tab:conf}, \ref{app:baselines}, \ref{app:samemin}, \ref{tab:grok}, .
This fragment-enabled reference requires fewer force calls on amino acid--ligand pairs and Solvated PubChem.
During evolution, both Astra and Fable enabled disconnected fragments and introduced further changes that reduced force-call cost while preserving mean energy recovery on the training and validation sets (Appendix~\ref{app:interpretation}).

\section{Autoresearch harness and loop}
\label{app:research_setup}

\subsection{Autoresearch harness}
\label{app:harness}

\paragraph{Isolated agent environments.}
Each search runs in an isolated environment to prevent access to other runs’ implementations, results, and research histories, preserving the independence of the experiments; web search remains allowed.
The container runs as an unprivileged user with a read-only root filesystem, read-only dataset mounts, and explicit CPU, memory, and process limits.
The same outer research loop supports three interchangeable agent interfaces: Claude Code, Codex CLI, and Cursor Agent.
The optimizer source, research instructions, training and validation splits, and acceptance protocol are initialized from the same snapshot across the three experiments.
The agent can develop and test candidate optimizers in its workspace, while formal scores are computed by the remote evaluator.
In our autoresearch experiments, Fable~5.1 (max) was run through Claude Code~2.1.251;
Astra (high) used Codex CLI~0.155.0;
Grok~4.6 (xhigh) used Cursor Agent~\texttt{2026.09.15-d2fe57e};

\paragraph{Evaluation relay and remote workers.}
The evaluator consists of a remote gateway, a shared Redis task queue, and a pool of workers distributed across CPU hosts.
Keeping this infrastructure outside the agent's environment protects the scoring procedure and reference data from modification, while distributing independent molecular optimizations allows each candidate to be evaluated in parallel across the dataset.
The agent submits its candidate through a fixed, run-specific SSH relay, which exposes the JSON evaluation protocol without granting general remote-shell access or exposing the host's SSH credentials.
The gateway associates the submission with its Git commit, source hash, dataset release, and training or validation split, then creates one optimization task per molecule.
Task identifiers enter a first-in, first-out Redis queue; task inputs, execution states, and results are stored separately under those identifiers.
Available workers claim tasks atomically from the shared queue and execute each complete molecular optimization in a guarded subprocess, obtaining energies and forces from GFN2-xTB.
The pool is comprised of 192 single-thread workers across 4 remote evaluation hosts with Intel Xeon Gold 6348 CPU.
Workers renew their task leases through periodic heartbeats and write force-call counts, energies, convergence status, and errors back to Redis.
The gateway collects these results into a durable evaluation record and returns aggregate metrics and per-molecule diagnostics to the agent.
Completed molecular results are retained when an evaluation is resumed; unresolved work leaves the evaluation pending rather than producing a score from a partial dataset.

A failure label or non-convergence flag gives the agent little information about what went wrong during the optimization.
The evaluator therefore supports diagnostic bundles for failed or non-converged optimizations, capturing the numerical evidence leading up to termination.
The bundle retains up to eight recent geometry/force frames, 200 scalar records of energies, timing, force-call counts and convergence metrics, the latest attempted geometry, and any available exception traceback tied to the candidate's source hash.
Bundles are size-limited and explicitly report missing or truncated evidence.
A separate collector stores the bundles, while Redis carries only compact descriptors.
Through the same restricted relay, the agent invokes the \texttt{diagnostics} command to list available bundles.
The agent can download a specific bundle into it's workspace after verifying its provenance and SHA-256 hash.
These commands retrieve existing evidence without launching calculations or changing evaluation scores and help the agent debug and improve next candidates.

\paragraph{Held-out calculations and DFT.}
The held-out GFN2-xTB evaluations use xTB~6.7.1 with accuracy parameter 0.2.
The Redis worker pool described above serves the xTB research evaluations.
The separate DFT evaluator runs the frozen optimizers on remote hosts, obtaining energies and analytical gradients from ORCA~6.1.1 through OPI at the \texttt{r2SCAN-3c} level with \texttt{TightSCF} and \texttt{ENGRAD} calculations.
Each force evaluation uses an isolated temporary directory and a configured ORCA core count.
Per-molecule records retain energies, force-call counts, and convergence status.

\subsection{The autoresearch loop}
\label{app:autoresearch}

This subsection describes in detail the rules and artifacts of the autoresearch process.

\paragraph{What the agent may touch.}
The following rules are an instruction for the agent, written in the \texttt{program.md}.
The agent edits and commits only \texttt{algo.py}, containing the optimizer, while maintaining separate research notes and saved implementations.
It is prohibited from modifying the evaluator, convergence criteria, evaluation clients, control scripts, molecule sets, or baseline metadata.
The convergence thresholds are disclosed, but the evaluator alone determines whether they are met.
Local work may include code editing, static checks, public-web research, and analysis of saved evaluation results; numerical optimizer experiments must use the remote evaluator.
The agent must not introduce molecule-specific recognition or special cases based on known identities, exact formulas, or identifying graph features.
The agent is also instructed to give a promising but failed idea up to three further, targeted repairs before reassessment.
This limit exists so the agent does not get stuck in hyperparameter tuning loop for an extended period of time.

\paragraph{Per-molecule statistics.}
The agent receives a JSON record for each training or validation evaluation, saved under \texttt{evaluation\_results/}.
Alongside aggregate metrics and evaluation identifiers, it contains per-molecule results with force-call counts and limits, relative force-call cost and energy recovery, final energies and geometries, convergence status, and stopping reasons.
Errors and failure details are recorded separately, distinguishing numerical failures from non-converged optimizations that return usable results.
The per-molecule records are the important part.
An aggregate tells the agent whether a mechanism helped; these records tell it \emph{which molecules} it helped and which it hurt, which is what turns the next cycle into a chemically-grounded hypothesis rather than a parameter sweep.

\paragraph{Agent's memory.}
Research artifacts persist across cycles and are consulted before choosing the next experiment.
The machine-written \texttt{results.tsv} records candidate and champion commits, evaluation identifiers, metrics, and decisions, including rejected candidates and pending evaluations.
Two derived tables, \texttt{generalizable.tsv} and \texttt{non\_generalizable.tsv}, retain the starting champion and subsequent accepted candidates, and candidates that failed the generalization gate, respectively.
The agent writes a chronological narrative in \texttt{full\_log.md}, describing each hypothesis, implementation change, interpretation, and next experiment.
The \texttt{backlog.md} file is a ledger of near-misses and promising future directions: each idea records its hypothesis, outcome and remaining uncertainty, a concrete reason to revisit it, and a proposed next experiment.
Promising rejected implementations are preserved under \texttt{ideas/}, indexed by idea identifier and full candidate commit hash, before the champion is restored.
Backlog entries link these implementations to their evaluation records and narrative history; subsequent attempts are appended without erasing earlier failures.
The agent updates each idea's status as deferred, revisiting, incorporated, or closed.

\paragraph{The cycle.}
The cycle is described in Algorithm~\ref{alg:ar}.
The agent first reads the decision tables, evaluation evidence, narrative log, and backlog.
The cycle manager allocates the next cycle or resumes an unfinished one, recording the champion and evaluation paths in \texttt{cycle.json}.
The agent implements one coherent hypothesis, commits \texttt{algo.py}, and evaluates the committed candidate on training through the remote client.
Only candidates passing the validity gate proceed to validation.
After the candidate is evaluated on validation set, the decision writer checks evaluation provenance and checks if is passes the generalization gate.
An accepted candidate replaces the champion.
For a rejected candidate, the agent first preserves any promising implementation and updates the backlog, then restores the champion while retaining research artifacts.
If evaluation is incomplete, the candidate remains unchanged and the same request is resumed before another candidate is started.
The agent puts a record to \texttt{full\_log.md} after each cycle and continues until the research run is stopped.

\FloatBarrier
\section{Per-molecule energy behaviour}
\label{app:dE}
\begin{table}[!htbp]
\centering
\caption{\textbf{Per-molecule energy difference}
$\Delta E = E^{\mathrm{AutoSella}}_{\mathrm{final}} - E^{\mathrm{Sella}}_{\mathrm{final}}$
\textbf{on xTB.} Negative is deeper than the reference, so every column is better when smaller.
\emph{Deeper} is the share reaching a strictly deeper minimum; the last two columns the shares finishing more than $0.1$ and
$1.0$\,kcal\,mol$^{-1}$ above it.
Statistics are over the molecules scored in Table~\ref{tab:main}, shares over the full benchmark $N$.}
\label{tab:dE}
\footnotesize
\setlength{\tabcolsep}{2.5pt}
\resizebox{\textwidth}{!}{%
\begin{tabular}{@{}ll S[table-format=-1.4, group-digits=none] S[table-format=-1.2e-1, group-digits=none, exponent-product=\times, retain-zero-exponent=true] S[table-format=2.2] S[table-format=-3.2] S[table-format=2.1] S[table-format=2.2] S[table-format=2.2]@{}}
\toprule
& & \multicolumn{4}{c}{$\Delta E$ (kcal\,mol$^{-1}$)\,$\downarrow$} & \multicolumn{3}{c}{share of molecules (\%)} \\
\cmidrule(lr){3-6}\cmidrule(l){7-9}
Benchmark & Method & {Mean} & {Median} & {Max} & {Min} & {Deeper\,$\uparrow$} & {$>0.1$\,$\downarrow$} & {$>1.0$\,$\downarrow$} \\
\midrule
\multirow{2}{*}{Test} & AutoSella-F & -0.0184 & -6e-06 & 3.77 & -7.69 & 74.8 & 1.00 & 0.40 \\
 & AutoSella-A & -0.0072 & -7e-06 & 3.75 & -6.91 & 73.4 & 0.60 & 0.20 \\
\cmidrule(l){2-9}
\multirow{2}{*}{Dipeptides} & AutoSella-F & -0.0422 & -8e-06 & 4.33 & -10.29 & 73.7 & 2.08 & 1.19 \\
 & AutoSella-A & -0.0179 & -5e-06 & 4.33 & -10.29 & 64.0 & 1.63 & 0.89 \\
\cmidrule(l){2-9}
\multirow{2}{*}{AA--ligand pairs} & AutoSella-F & -0.2982 & -5e-06 & 32.24 & -19.32 & 60.6 & 26.95 & 18.74 \\
 & AutoSella-A & -0.0673 & -4e-06 & 25.13 & -21.12 & 67.4 & 12.84 & 9.05 \\
\cmidrule(l){2-9}
\multirow{2}{*}{Solvated PubChem} & AutoSella-F & -2.6056 & -1.3 & 20.89 & -103.73 & 57.0 & 41.00 & 38.00 \\
 & AutoSella-A & 0.4837 & -3e-06 & 30.37 & -33.93 & 49.0 & 40.00 & 36.00 \\
\bottomrule
\end{tabular}%
}
\end{table}

Table~\ref{tab:dE} decomposes the aggregate energy ratio of Table~\ref{tab:main} into the distribution behind it.
On the single-molecule Test and Dipeptides datasets, the median difference from the reference ranges from $-8\times10^{-6}$ to $-5\times10^{-6}$\,kcal\,mol$^{-1}$: on a typical molecule, both evolved optimizers reach essentially the reference energy.
Only between $0.60$ and $2.08\%$ molecules end more than $0.1$\,kcal\,mol$^{-1}$ above it and between $0.20$ and $1.19\%$ more than $1.0$\,kcal\,mol$^{-1}$ above, making the
higher-energy endpoints rather rare.

For systems with fragments, the energy differences are larger.
AutoSella-F reaches final energies more than $1.0$\,kcal\,mol$^{-1}$ below Sella on $24.84\%$ of AA--ligand pairs and $51.00\%$ of Solvated PubChem systems.
Its median difference remains tiny on AA--ligand pairs ($-5\times10^{-6}$\,kcal\,mol$^{-1}$), but shifts to $-1.30$\,kcal\,mol$^{-1}$ on Solvated PubChem; the corresponding means are $-0.2982$ and $-2.6056$\,kcal\,mol$^{-1}$.
These gains are not uniform: AutoSella-F also finishes more than $1.0$\,kcal\,mol$^{-1}$ above the reference on $18.74\%$ and $38.00\%$ of these datasets, respectively.
For AutoSella-A, the corresponding higher-energy shares are $9.05\%$ and $36.00\%$.
On Solvated PubChem, its mean difference is positive ($0.4837$\,kcal\,mol$^{-1}$), despite a nearly zero median ($-3\times10^{-6}$\,kcal\,mol$^{-1}$).

A lower final energy alone does not establish that an optimizer has reached a deeper local minimum, which is why the structural and frequency checks of Appendix~\ref{app:conf} are reported.

\section{Structural agreement with the reference}
\label{app:conf}
\begin{table}[!htbp]
\centering
\caption{\textbf{Do the optimizers reach the same
local minimum as Sella?} All results on xTB.
Each molecule is checked three ways: heavy-atom RMSD to the Sella geometry after optimal superposition, $|\Delta E|$ against the Sella final energy, and the vibrational spectrum from
a numerical Hessian.
The four $\nu$ columns count molecules carrying at least one
mode below the stated cutoff, which separate finite-difference noise and near-free torsions from a genuine saddle.
\emph{Same min.}\ requires RMSD $<0.1$\,\AA, $|\Delta E| < 0.1$\,kcal\,mol$^{-1}$ and no mode below $-50$\,cm$^{-1}$, all at once.
RMSD and $|\Delta E|$ are undefined for the reference against itself; its $\nu$ counts are given because it is not always at a true minimum either.}
\label{tab:conf}
\footnotesize
\setlength{\tabcolsep}{2pt}
\resizebox{\textwidth}{!}{%
\begin{tabular}{@{}llr S[table-format=1.4] S[table-format=2.1] S[table-format=2.1] rrrr S[table-format=2.1]@{}}
\toprule
& & & \multicolumn{2}{c}{RMSD to Sella} & & \multicolumn{4}{c}{molecules with $\nu <$ (cm$^{-1}$)\,$\downarrow$} & \\
\cmidrule(lr){4-5}\cmidrule(lr){7-10}
Benchmark & Method & {$N$} & {med (\AA)\,$\downarrow$} & {$<0.1$\,\AA\,\%\,$\uparrow$} & {$|\Delta E|_{<0.1}$\,\%\,$\uparrow$} & {$0$} & {$-50$} & {$-200$} & {$-300$} & {Same min.\,\%\,$\uparrow$} \\
\midrule
\multirow{3}{*}{Test} & AutoSella-F & \multirow{3}{*}{500} & 0.0005 & 97.6 & 97.6 & 2 & 0 & 0 & 0 & 97.2 \\
 & AutoSella-A &  & 0.0005 & 97.8 & 98.2 & 4 & 1 & 1 & 1 & 97.6 \\
 & Sella (ref.) &  & {--} & {--} & {--} & 4 & 1 & 0 & 0 & {--} \\
\cmidrule(l){2-11}
\multirow{3}{*}{Dipeptides} & AutoSella-F & \multirow{3}{*}{673} & 0.0008 & 93.6 & 93.9 & 2 & 1 & 0 & 0 & 93.2 \\
 & AutoSella-A &  & 0.0008 & 95.8 & 96.3 & 2 & 0 & 0 & 0 & 95.7 \\
 & Sella (ref.) &  & {--} & {--} & {--} & 1 & 1 & 0 & 0 & {--} \\
\cmidrule(l){2-11}
\multirow{3}{*}{AA--ligand pairs} & AutoSella-F & \multirow{3}{*}{475} & 1.1763 & 40.0 & 42.3 & 5 & 1 & 0 & 0 & 39.8 \\
 & AutoSella-A &  & 0.0014 & 69.7 & 71.4 & 3 & 0 & 0 & 0 & 69.5 \\
 & Sella (ref.) &  & {--} & {--} & {--} & 6 & 0 & 0 & 0 & {--} \\
\cmidrule(l){2-11}
\multirow{3}{*}{Solvated PubChem} & AutoSella-F & \multirow{3}{*}{100} & 2.0068 & 1.0 & 1.0 & 0 & 0 & 0 & 0 & 1.0 \\
 & AutoSella-A &  & 1.0807 & 18.0 & 19.0 & 0 & 0 & 0 & 0 & 18.0 \\
 & Sella (ref.) &  & {--} & {--} & {--} & 0 & 0 & 0 & 0 & {--} \\
\bottomrule
\end{tabular}%
}
\end{table}

The validity gate compares energies, so an optimizer could in principle pass it while settling in a different well from the reference.
Table~\ref{tab:conf} tests that directly, on GFN2-xTB, for all four held-out datasets.
Each molecule is checked three independent ways against the Sella result: heavy-atom RMSD after optimal superposition, the absolute energy difference $|\Delta E|$, and the vibrational spectrum from a numerical Hessian.
RMSD is computed after optimal translation and rotation, using fixed atom indices and heavy atoms; systems with fewer than three heavy atoms use all atoms.
The spectrum is summarized by the number of molecules with at least one frequency below each reported cutoff. A molecule satisfies our “same minimum” criterion when RMSD \(<0.1\) Å, \(|\Delta E|<0.1\) kcal mol\(^{-1}\), and the candidate spectrum contains no frequency below \(-50\) cm\(^{-1}\). This cutoff provides a practical tolerance for the comparison rather than proof of a local minimum.

AutoSella-F and AutoSella-A satisfy the same-minimum criterion on $97.2\%$ and $97.6\%$ of Test molecules, and $93.2\%$ and $95.7\%$ of Dipeptides, respectively.
The median RMSD over jointly converged pairs is $5\times10^{-4}$\,\AA{} on Test and $8\times10^{-4}$\,\AA{} on Dipeptides for both optimizers, more than two orders of magnitude below the $0.1$\,\AA{} threshold.
Agreement falls to $39.8\%$ and $69.5\%$ on AA--ligand pairs, and to $1.0\%$ and $18.0\%$ on Solvated PubChem.
We hypothesize this is due to a greater variety of relative fragment poses and intermolecular contacts, including alternative arrangements of the surrounding water molecules in Solvated PubChem.
Low agreement with Sella, however, need not imply poor relaxation: the alternative endpoints on average lie below the reference energy, as Table~\ref{tab:dE} shows.

Imaginary modes are rare for every method, the reference included: on Test, Sella carries a mode below zero on four molecules, against two for AutoSella-F and four for AutoSella-A.
No negative frequencies are reported in any Solvated PubChem spectra.
Across all four datasets, every pair satisfying both the geometry and energy thresholds also passes the candidate frequency cutoff.
The lower agreement on fragment systems therefore reflects differences in final geometry and energy, rather than rejection by the frequency screen.

%
%

\section{Optimizers}
\label{app:panel}
\label{sec:classical-baselines}

Comparing against Sella alone establishes that the evolved programs improve on the algorithm they started from, but not that Sella~2.5.0 is itself a strong point of reference.
This section measures a panel of established open-source optimizers under the identical evaluation protocol, so the results in Table~\ref{tab:main} and Table~\ref{tab:eval} can be placed on an absolute scale.

\subsection{The classical panel}
\label{sec:classical-panel}

\begin{table}[!t]
\centering
\caption{\textbf{Every classical configuration measured, relative to Sella~2.5.0 on GFN2-xTB.} For each benchmark set: $C$, the per-molecule cost $C_{\text{per-mol}}$ of Eq.~\eqref{eq:cost}, $\bar r$ (Eq.~\eqref{eq:energy}) and the number of molecules without a converged result (Err.). $C$ and $\bar r$ are means over the molecules that both the reference and the method converged
; ``--'' means that no molecule converged. Every optimizer uses the same harness, backend, force calls limit and convergence criteria, with its own stopping test disabled. Best values among non-Sella competitors in each column, including ties at the displayed precision, are bold.}
\label{app:baselines}
\footnotesize
\setlength{\tabcolsep}{0.9pt}
\begin{tabular}{@{}l S[table-format=3.1] S[table-format=1.4] S[table-format=3.0] S[table-format=3.1] S[table-format=1.4] S[table-format=3.0] S[table-format=3.1] S[table-format=1.4] S[table-format=3.0] S[table-format=3.1] S[table-format=1.4] S[table-format=3.0]@{}}
\toprule
& \multicolumn{3}{c}{Test} & \multicolumn{3}{c}{Dipeptides} & \multicolumn{3}{c}{AA--ligand} & \multicolumn{3}{c}{Solvated PubChem} \\
& \multicolumn{3}{c}{$N=500$} & \multicolumn{3}{c}{$N=673$} & \multicolumn{3}{c}{$N=475$} & \multicolumn{3}{c}{$N=100$} \\
\cmidrule(lr){2-4}\cmidrule(lr){5-7}\cmidrule(lr){8-10}\cmidrule(l){11-13}
Method & {$C$\,$\downarrow$} & {$\bar r$\,$\uparrow$} & {Err.} & {$C$\,$\downarrow$} & {$\bar r$\,$\uparrow$} & {Err.} & {$C$\,$\downarrow$} & {$\bar r$\,$\uparrow$} & {Err.} & {$C$\,$\downarrow$} & {$\bar r$\,$\uparrow$} & {Err.} \\
\midrule
Sella / internal \emph{(ref.)} & 100.0 & 1.0000 & 0 & 100.0 & 1.0000 & 0 & 100.0 & 1.0000 & 0 & 100.0 & 1.0000 & 2 \\
\midrule
pysis.\ / DLC & \bfseries 105.4 & 0.9999 & 8 & \bfseries 103.6 & 0.9997 & 6 & 152.6 & 0.9346 & 99 & 198.6 & 0.9557 & 92 \\
pysis.\ / redund., Fischer & 110.5 & 0.9999 & 34 & 106.0 & \bfseries 1.0009 & \bfseries 0 & 129.7 & 0.9175 & 191 & {--} & {--} & 100 \\
pysis.\ / TRIC & 110.7 & 0.9993 & 34 & 105.2 & \bfseries 1.0009 & \bfseries 0 & \bfseries 120.4 & 1.0044 & \bfseries 11 & \bfseries 130.8 & 0.9917 & 80 \\
pysis.\ / redund., Lindh & 110.8 & 1.0004 & 42 & 112.1 & 0.9998 & 1 & 140.4 & 0.9011 & 204 & {--} & {--} & 100 \\
geomeTRIC / TRIC & 129.6 & 1.0011 & \bfseries 0 & 141.1 & 1.0005 & \bfseries 0 & 196.2 & 1.0082 & 18 & 133.1 & \bfseries 1.0014 & \bfseries 1 \\
geomeTRIC / DLC & 130.4 & 1.0011 & \bfseries 0 & 141.5 & 1.0005 & \bfseries 0 & 176.0 & 1.0180 & 12 & 219.2 & 0.9926 & 54 \\
pysis.\ / redund., Swart & 130.5 & 0.9999 & 47 & 125.7 & 1.0005 & 6 & 143.8 & 0.9230 & 173 & {--} & {--} & 100 \\
geomeTRIC / primitive & 145.9 & \bfseries 1.0016 & \bfseries 0 & 154.2 & 0.9998 & 2 & 183.1 & \bfseries 1.0190 & 18 & 234.7 & 0.9868 & 69 \\
pysis.\ / redund., simple & 193.2 & 1.0013 & 43 & 201.3 & 0.9999 & 4 & 193.3 & 0.9114 & 190 & {--} & {--} & 100 \\
PyBerny & 204.3 & 0.9986 & 56 & 250.9 & 0.9946 & 154 & 310.2 & 0.8579 & 404 & {--} & {--} & 100 \\
geomeTRIC / HDLC & 327.2 & 1.0002 & 21 & 351.5 & 0.9994 & 48 & 378.0 & 0.9294 & 380 & 213.1 & 0.9786 & 56 \\
pysis.\ / redund., unit & 444.9 & 0.9986 & 56 & 455.3 & 0.9983 & 128 & 312.7 & 0.8002 & 338 & {--} & {--} & 100 \\
ASE PreconLBFGS & 455.0 & 0.9988 & 43 & 443.4 & 0.9978 & 149 & 383.4 & 0.9058 & 420 & 261.2 & 0.9816 & 76 \\
geomeTRIC / Cartesian & 572.2 & 0.9967 & 141 & 533.9 & 0.9938 & 444 & 367.6 & 0.7739 & 458 & 287.2 & 0.9811 & 95 \\
Sella / Cartesian & 598.7 & 1.0005 & 106 & 548.2 & 0.9991 & 352 & 499.0 & 0.9439 & 465 & 282.9 & 0.9799 & 87 \\
ASE LBFGS & 606.2 & 0.9961 & 181 & 561.8 & 0.9916 & 518 & 340.1 & 0.6751 & 457 & 345.5 & 0.9824 & 95 \\
ASE BFGS & 606.7 & 0.9961 & 179 & 547.6 & 0.9926 & 519 & 317.9 & 0.6830 & 460 & 324.0 & 0.9773 & 94 \\
SciPy BFGS & 608.0 & 0.9960 & 183 & 548.8 & 0.9925 & 524 & 358.6 & 0.6988 & 461 & 331.8 & 0.9865 & 96 \\
ASE FIRE2 & 642.1 & 0.9679 & 262 & 467.7 & 0.9001 & 550 & 236.8 & 0.5301 & 300 & {--} & {--} & 100 \\
ASE LBFGSLineSearch & 645.2 & 0.9933 & 228 & 567.0 & 0.9827 & 592 & 285.2 & 0.5804 & 462 & 304.5 & 0.9315 & 99 \\
pysis.\ / Cart., unit & 646.0 & 0.9949 & 207 & 560.1 & 0.9881 & 577 & 327.2 & 0.6407 & 463 & 330.4 & 0.9891 & 96 \\
ASE BFGSLineSearch & 666.6 & 0.9973 & 205 & 592.4 & 0.9978 & 552 & 448.4 & 0.7732 & 472 & 273.7 & 0.9833 & 92 \\
ASE FIRE & 689.4 & 0.9638 & 206 & 478.2 & 0.9054 & 501 & 247.5 & 0.5321 & 279 & {--} & {--} & 100 \\
SciPy CG & 699.5 & 0.9865 & 220 & 540.4 & 0.9482 & 524 & 326.1 & 0.6032 & 388 & 317.9 & 0.9315 & 99 \\
ASE ODE12r & 737.7 & 0.9688 & 274 & 523.6 & 0.8971 & 590 & 264.9 & 0.5246 & 335 & {--} & {--} & 100 \\
ASE MDMin & 935.0 & 0.9983 & 489 & {--} & {--} & 673 & {--} & {--} & 475 & {--} & {--} & 100 \\
\bottomrule
\end{tabular}
\end{table}

\begin{table}[!t]
\centering
\caption{\textbf{Do the classical optimizers reach the same local minimum as Sella?} Criteria as in Table~\ref{tab:conf}; only molecules that both the method and Sella converged can count. Three columns per benchmark: \emph{\%} reaching the same minimum, as a share of $N$; \emph{\AA}, the median heavy-atom RMSD to the Sella geometry; and \emph{$\nu$}, the number of molecules with a mode below $-50$\,cm$^{-1}$.}
\label{app:samemin}
\footnotesize
\setlength{\tabcolsep}{1.15pt}
\begin{tabular}{@{}l S[table-format=3.1] S[table-format=1.4] S[table-format=3.0] S[table-format=3.1] S[table-format=1.4] S[table-format=3.0] S[table-format=3.1] S[table-format=1.4] S[table-format=3.0] S[table-format=3.1] S[table-format=1.4] S[table-format=3.0]@{}}
\toprule
& \multicolumn{3}{c}{Test} & \multicolumn{3}{c}{Dipeptides} & \multicolumn{3}{c}{AA--ligand} & \multicolumn{3}{c}{Solvated PubChem} \\
& \multicolumn{3}{c}{$N=500$} & \multicolumn{3}{c}{$N=673$} & \multicolumn{3}{c}{$N=475$} & \multicolumn{3}{c}{$N=100$} \\
\cmidrule(lr){2-4}\cmidrule(lr){5-7}\cmidrule(lr){8-10}\cmidrule(l){11-13}
Method & {\%\,$\uparrow$} & {\AA\,$\downarrow$} & {$\nu$\,$\downarrow$} & {\%\,$\uparrow$} & {\AA\,$\downarrow$} & {$\nu$\,$\downarrow$} & {\%\,$\uparrow$} & {\AA\,$\downarrow$} & {$\nu$\,$\downarrow$} & {\%\,$\uparrow$} & {\AA\,$\downarrow$} & {$\nu$\,$\downarrow$} \\
\midrule
Sella / internal \emph{(ref.)} & {--} & {--} & 1 & {--} & {--} & 1 & {--} & {--} & 0 & {--} & {--} & 0 \\
\midrule
pysis.\ / DLC & \bfseries 97.0 & \bfseries 0.0004 & 1 & 95.1 & \bfseries 0.0006 & 1 & 16.4 & 1.6320 & 1 & 0.0 & 2.1871 & \bfseries 0 \\
pysis.\ / redund., Fischer & 92.4 & \bfseries 0.0004 & \bfseries 0 & 95.7 & \bfseries 0.0006 & 1 & 13.3 & 1.7364 & \bfseries 0 & 0.0 & {--} & {--} \\
pysis.\ / TRIC & 92.4 & \bfseries 0.0004 & \bfseries 0 & 95.8 & \bfseries 0.0006 & \bfseries 0 & \bfseries 63.4 & \bfseries 0.0020 & \bfseries 0 & 2.0 & 1.3388 & \bfseries 0 \\
pysis.\ / redund., Lindh & 89.6 & 0.0005 & \bfseries 0 & \bfseries 96.7 & 0.0008 & \bfseries 0 & 13.7 & 1.6537 & \bfseries 0 & 0.0 & {--} & {--} \\
geomeTRIC / TRIC & 96.8 & 0.0007 & 3 & 95.4 & 0.0012 & 2 & 45.5 & 0.5352 & \bfseries 0 & \bfseries 3.0 & 1.4768 & \bfseries 0 \\
geomeTRIC / DLC & 96.8 & 0.0007 & 3 & 95.5 & 0.0012 & 2 & 38.5 & 1.1446 & \bfseries 0 & 0.0 & 2.0138 & \bfseries 0 \\
pysis.\ / redund., Swart & 88.6 & 0.0005 & \bfseries 0 & 93.9 & 0.0009 & 1 & 14.3 & 1.7165 & \bfseries 0 & 0.0 & {--} & {--} \\
geomeTRIC / primitive & 95.2 & 0.0008 & 3 & 95.5 & 0.0013 & 2 & 38.7 & 1.0453 & \bfseries 0 & 0.0 & 1.9521 & \bfseries 0 \\
pysis.\ / redund., simple & 86.4 & 0.0018 & 3 & 94.8 & 0.0029 & 4 & 12.6 & 1.6245 & 1 & 0.0 & {--} & {--} \\
PyBerny & 78.6 & 0.0010 & 11 & 62.0 & 0.0012 & 23 & 5.3 & 1.2918 & 1 & 0.0 & {--} & {--} \\
geomeTRIC / HDLC & 79.6 & 0.0059 & 2 & 78.8 & 0.0100 & 1 & 8.4 & 0.3811 & \bfseries 0 & 0.0 & 1.6091 & 1 \\
pysis.\ / redund., unit & 79.0 & 0.0079 & 9 & 73.1 & 0.0120 & 14 & 3.6 & 1.6415 & 4 & 0.0 & {--} & {--} \\
ASE PreconLBFGS & 73.0 & 0.0177 & 8 & 60.3 & 0.0266 & 13 & 3.6 & 0.6273 & \bfseries 0 & 0.0 & 1.5516 & \bfseries 0 \\
geomeTRIC / Cartesian & 52.4 & 0.0286 & 5 & 21.0 & 0.0680 & 4 & 0.4 & 1.9348 & \bfseries 0 & 0.0 & 1.2047 & \bfseries 0 \\
Sella / Cartesian & 67.6 & 0.0059 & 3 & 42.1 & 0.0134 & \bfseries 0 & 1.1 & 0.2221 & \bfseries 0 & 0.0 & 1.3288 & \bfseries 0 \\
ASE LBFGS & 45.6 & 0.0292 & 7 & 13.4 & 0.0708 & 5 & 0.0 & 2.1225 & \bfseries 0 & 0.0 & \bfseries 1.1585 & \bfseries 0 \\
ASE BFGS & 45.8 & 0.0305 & 7 & 13.1 & 0.0756 & 5 & 0.0 & 1.9407 & \bfseries 0 & 0.0 & 1.1825 & \bfseries 0 \\
SciPy BFGS & 45.4 & 0.0278 & 7 & 12.6 & 0.0801 & 6 & 0.0 & 1.9609 & \bfseries 0 & 0.0 & 1.1690 & \bfseries 0 \\
ASE FIRE2 & 9.0 & 0.1898 & 18 & 0.0 & 0.4886 & 60 & 0.0 & 2.0425 & 69 & 0.0 & {--} & {--} \\
ASE LBFGSLineSearch & 35.6 & 0.0384 & 10 & 5.8 & 0.0852 & 4 & 0.0 & 2.5854 & \bfseries 0 & 0.0 & 1.5538 & \bfseries 0 \\
pysis.\ / Cart., unit & 40.8 & 0.0338 & 10 & 7.7 & 0.0841 & 4 & 0.0 & 2.2209 & \bfseries 0 & 0.0 & 1.1734 & \bfseries 0 \\
ASE BFGSLineSearch & 46.6 & 0.0136 & 3 & 14.1 & 0.0320 & \bfseries 0 & 0.2 & 0.5743 & \bfseries 0 & 0.0 & 1.3342 & \bfseries 0 \\
ASE FIRE & 9.4 & 0.2207 & 21 & 0.0 & 0.4836 & 76 & 0.0 & 2.0375 & 71 & 0.0 & {--} & {--} \\
SciPy CG & 22.0 & 0.1023 & 12 & 0.0 & 0.4305 & 29 & 0.0 & 2.1003 & 9 & 0.0 & 1.5629 & \bfseries 0 \\
ASE ODE12r & 8.0 & 0.1628 & 17 & 0.0 & 0.4598 & 44 & 0.0 & 2.0494 & 57 & 0.0 & {--} & {--} \\
ASE MDMin & 1.6 & 0.0098 & \bfseries 0 & 0.0 & {--} & {--} & 0.0 & {--} & {--} & 0.0 & {--} & {--} \\
\bottomrule
\end{tabular}
\end{table}

\paragraph{The evaluation protocol.}
Every optimizer is driven through the same interface, the same GFN2-xTB backend~\citep{gfn2xtb2019,xtbreview2021}, the same budget of force calls (200, and 500 on Solvated PubChem, as the benchmark specifies), and the same five Gaussian-style convergence criteria, on all four benchmark sets.
Each optimizer's own stopping test is disabled.
One optimizer required non-default settings forced by this system class rather than chosen: PreconLBFGS uses the exponential preconditioner with explicit $r_{\text{NN}} = 1.5$\,\AA{} and $r_{\text{cut}} = 3.0$\,\AA{}, because the automatic neighbour-cutoff estimate does not terminate for an isolated molecule in no cell; the two radii bracket a covalent bond and its second shell, the range the preconditioner is built for.

Table~\ref{app:baselines} reports force-call cost $C_{\text{per-mol}}$ and $\bar r$ for the reference and all 26 classical configurations on each of the four sets, in the units of Table~\ref{tab:main}, and Table~\ref{app:samemin} applies the structural criteria of Appendix~\ref{app:conf} to every one of them.
Sella has the lowest per-molecule force-call cost among the tested classical optimizers on all four datasets, supporting its selection as the starting point for evolution.

Our experiments show that the choice of a coordinate system plays a crucial role in the optimization process. 
On Dipeptides, pysisyphus in TRIC coordinates reaches the same minimum as the reference on $95.8\%$ of molecules and geomeTRIC in TRIC on $95.4\%$, while Cartesian BFGS manages $23.3\%$ and FIRE reaches it on
\emph{none} of the 673 --- its median heavy-atom RMSD there is $0.56$\,\AA, not a marginally different conformer but a different structure.
This is the mechanism behind the cost column: Cartesian methods are not merely slower, they frequently stop somewhere else.

\subsection{What actually makes an optimizer efficient}
\label{sec:two-factors}

Shajan et al.~\citep{shajan2023} identify two factors behind the efficiency of a geometry optimizer: the coordinate system in which it takes its steps, and the initial guess for the Hessian. 
On the Baker set~\citep{baker1993} they find $20$--$24$ steps per molecule in Cartesian coordinates with a unit initial Hessian, $11$--$12$ in internal coordinates with a unit Hessian, and $6$--$10$ in internal coordinates with a model Hessian. 
Each of these results, however, comes from a different program, so the two factors cannot be separated from each other or from differences in implementation, as the authors note.
Our panel allows a direct test: we change one factor at a time inside a single program and keep everything else fixed.

\emph{Coordinate system.} In geomeTRIC~\citep{geometric2016} we keep the step algorithm, trust radius and Hessian guess the same and change only the coordinates.
Relative to Sella, the cost on Test rises from $129.6\%$ with TRIC and $130.4\%$ with delocalized internals~\citep{bakerdlc1996}, through $145.9\%$ with primitive internals and $327.2\%$ with HDLC~\citep{hdlc2000}, to $572.2\%$ with Cartesian coordinates: the coordinate system alone changes the cost by a factor of $4.4$.

\emph{Initial Hessian.} In pysisyphus we keep redundant internal coordinates and change only the initial Hessian. The cost is $110.5\%$ with the Fischer--Alml\"of model~\citep{fischer1992}, $110.8\%$ with Lindh's~\citep{lindh1995}, $130.5\%$ with Swart's~\citep{swart2006}, $193.2\%$ with a simple diagonal guess and $444.9\%$ with the unit matrix: the initial Hessian alone changes the cost by a factor of $4.0$.

Both factors therefore matter, and by a similar amount.
These factors are averaged over the molecules that both the method and Sella converged, so they are only lower bounds: the worst settings also fail on more molecules --- geomeTRIC with Cartesian coordinates leaves 141 of the 500 Test molecules unconverged, and pysisyphus with the unit Hessian 56.

The consequence for the present work is that, on single-fragment molecules, Sella's margin over a well-configured generic optimizer is small. 
pysisyphus with delocalized internal coordinates needs $105.4\%$ of Sella's force calls on Test and $103.6\%$ on Dipeptides; with redundant internals and the Fischer--Alml\"of initial Hessian --- the same model Hessian Sella uses~\citep{sella2021geodesic} --- it needs $110.5\%$ and $106.0\%$. These averages are over molecules both optimizers converged; pysisyphus also fails on 8 (delocalized) and 34 (redundant) of the 500 Test molecules, all of which Sella converges. 
Almost all of the $5$--$6.5\times$ advantage that Sella holds over Cartesian methods therefore comes from its choice of internal coordinates and a model initial Hessian, rather than from its step algorithm.


\subsection{The optimizers}
\label{sec:optimizer-list}

The evolved optimizers, AutoSella-F and AutoSella-A, are described in Section~\ref{sec:gates} and Appendix~\ref{app:autoresearch}.
This section describes the classical optimizers of the panel, all run under the same force-call budget and convergence test.

\paragraph{Sella}~\citep{sella2022} performs restricted-step partitioned rational-function optimization in redundant internal coordinates, with iterative partial Hessian diagonalization, TS-BFGS-style Hessian updates, automatic dummy-atom handling for linear bends, and geodesic stepping on the internal-coordinate manifold~\citep{sella2021geodesic}.
It is the reference throughout this work. 
Configured in Cartesian coordinates it costs $5$--$6\times$ more, which is the single largest configuration effect we measured.

\paragraph{geomeTRIC}~\citep{geometric2016} introduces translation--rotation internal coordinates (TRIC), which add the translations and rotations of each molecular fragment to the primitive internal set.
It uses BFGS with a Schlegel-model initial Hessian~\citep{schlegel1984} and supports Cartesian, primitive, delocalized~\citep{bakerdlc1996} and hybrid-delocalized~\citep{hdlc2000} coordinates.
It is second on step count in Shajan et al.~\citep{shajan2023}.
Here it trails pysisyphus on single-fragment molecules, but its TRIC coordinates make it the most reliable classical optimizer on the multi-fragment sets: it converges on 99 of the 100 solvated systems.

\paragraph{pysisyphus}~\citep{pysisyphus2021} is a Python toolkit for exploring potential energy surfaces: rational-function optimization~\citep{rfo1985}, chain-of-states methods, and IRC.
It exposes the coordinate system and the model initial Hessian as independent options --- Fischer--Alml\"of~\citep{fischer1992}, Lindh~\citep{lindh1995}, Swart~\citep{swart2006}, simple, or unit --- and is consequently the only package in the panel that can separate the two factors.
With delocalized internal coordinates it is the strongest classical optimizer on single-fragment molecules, at  $105\%$ of Sella's force calls on Test and $104\%$ on Dipeptides; its variants that bridge fragments with artificial bonds, however, fail on most multi-fragment systems.

\paragraph{PyBerny}~\citep{pyberny} is an open implementation of the Berny algorithm used in Gaussian: quasi-Newton steps in redundant internal coordinates with an iterative Hessian estimate, a trust region, line search and coordinate weighting, accelerated by GDIIS~\citep{gdiis1984} and its energy-represented variant GEDIIS~\citep{gediis2006}. 
It ties with Sella for fewest steps in Shajan et al.~\citep{shajan2023}, but under our tighter five-criterion test it costs $2.0\times$ on Test and $2.5\times$ on Dipeptides, and reaches the reference minimum on only $62.7\%$ of Dipeptides.
On the solvated systems it cannot build its internal coordinates for 63 of 100 molecules.

\paragraph{ASE BFGS, LBFGS and their line-search variants}~\citep{ase2017} are
Cartesian quasi-Newton methods; LBFGS uses the limited-memory update~\citep{nocedal1980,lbfgs1989}.
Their initial inverse Hessian is a scaled identity ($\alpha = 70$ by default).
They cost $5.5$--$6.7\times$ the reference.
BFGSLineSearch is reported elsewhere as the most reliable optimizer for landing in the correct basin; here it is the most reliable of the Cartesian quasi-Newton group on energy fidelity while also the most expensive of them.

\paragraph{FIRE}~\citep{fire2006} and \textbf{FIRE2}~\citep{fire2020} are damped molecular-dynamics minimizers with adaptive time-stepping, widely used for their robustness and matrix-free cost per step.
On drug-like molecules under a 200-call budget they under-relax: median energy recovery falls to $0.92$--$0.93$ of the reference on Dipeptides, and neither reaches the reference minimum on any dipeptide.

\paragraph{SciPy CG and BFGS}~\citep{scipy2020} are the Polak--Ribi\`ere conjugate-gradient and BFGS implementations wrapped by ASE.
SciPy BFGS tracks ASE BFGS closely, as expected; conjugate gradient is consistently worse on energy fidelity.

\paragraph{ODE12r}~\citep{ode12r2019} is an adaptive-timestep ODE solver originally developed for minimum-energy-path finding, applied here as a minimizer. Like FIRE it under-relaxes on flexible molecules.

\paragraph{PreconLBFGS}~\citep{precon2016} applies a preconditioner to L-BFGS; the \texttt{Exp} preconditioner used here is built from interatomic distances.
It is the best-performing Cartesian-space method in the panel at $4.4$--$4.6\times$, consistent with the original report that preconditioning helps most for large systems --- our molecules, at 3--110 atoms, are below the regime where the published order-of-magnitude gains appear.

\section{Optimizer components and ablation reproducibility}
\label{app:ablations}
\label{app:interpretation}

This section interprets the improvements proposed by autoresearch (AR) by grouping the evolved mechanisms into functional blocks and measuring the effect of removing them.
We compare against the Sella with \texttt{allow\_fragments=false} used at the beginning of evolution.
The blocks summarize the implemented improvements related to a distinct mechanism in the optimizer (e.g. Hessian initialization or step size control).

\subsection{Experimental design}
Each single-block ablation starts from AutoSella-A/F and disables one or two family of changes. It therefore measures that family's effect conditional on the remaining evolved mechanisms. All variants are evaluated on both training and validation splits. The compared metric is $C_{\text{per-mol}}$ from Eq.~\ref{eq:cost}, the arithmetic mean of per-molecule force-call ratios. A valid evaluation requires complete coverage, no optimizer errors, and mean normalized energy recovery $\bar r\geq1-10^{-9}$ (Eq.~\ref{eq:energy}).

\begin{figure}[!h]
\centering
\includegraphics[width=\linewidth]{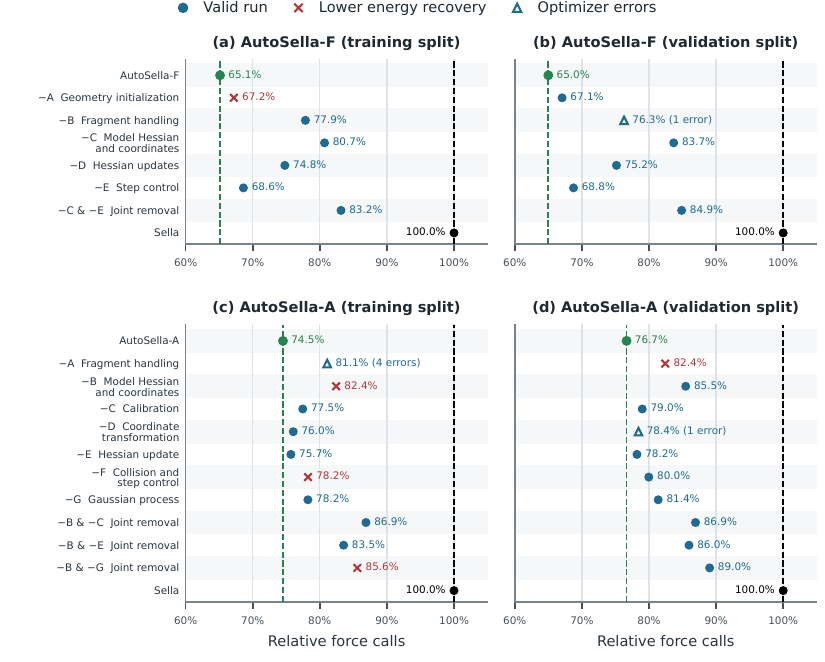}
\caption{\textbf{Ablation study on AutoSella improvements} Blocks are removed from AutoSella optimizers and their performance is compared with the Sella used at the beginning of evolution.
Relative force calls are defined in Eq.~\ref{eq:cost}.
Green circles mark the champions; black circles mark Sella.
Other valid results are blue.
A minus sign denotes removal; an ampersand denotes joint removal.
Red crosses mark mean $\bar r < 1$ (Eq.~\ref{eq:energy}).
For variants with optimizer errors, open triangles show means over successful cases only.}
\label{fig:family-ablations}
\end{figure}

\subsection{AutoSella-F blocks}

\paragraph{A: Geometry initialization.}
Before the first call to the potential, AutoSella-F adjusts the input geometry in three ways. None of them calls the potential, so they cost nothing in the force-call budget.

\emph{Geometric rules.} Some groups in the input start in a chemically known unfavourable conformation and are moved to a favourable one by a fixed geometry-based rule that depends only on bonding pattern.
A methyl group whose C--H bonds start lined up with the bonds of the neighbouring atom, which is the top of its rotation barrier, is rotated so that they lie in between those bonds instead; next to a C=O group the preferred orientation is the opposite, with one C--H lined up with the C=O bond.
An atom that is normally pyramidal, such as three-coordinate phosphorus, is bent out of plane when it starts flat. Most rules reflect standard conformational preferences.

\emph{Fragment pre-optimization.}
In a system of several molecules, the molecules are first moved and rotated as rigid bodies to minimize a simple model of their electrostatic and van der Waals interaction.
Partial charges come from formal charges and bond polarities, and the van der Waals terms use UFF parameters~\citep{uff1992}.
This brings the molecules close to a favourable mutual arrangement before the first call to the potential.

\emph{Symmetry breaking.}
Finally, in a system of several fragments, each fragment is shifted as a whole by 0.02\,\AA{} in a fixed pseudo-random direction, so that the optimizer does not stall at a point that is stationary only because of an exact symmetry of the start.

The ablation removes all three.

\paragraph{B: Fragment handling.} \label{par:f-fragments}
When a system consists of several disconnected fragments, Sella can describe each fragment by its own position and orientation in addition to its internal coordinates, so that fragments move relative to one another as rigid units.
Without this option, Sella instead links the fragments by artificial bonds that exist only in its coordinate system, not in the potential.
The starting program of the evolution did not use the option, and AutoSella-F enabled it early on; the reference Sella in this work uses it as well, so this block is shared with the reference rather than an advantage over it.
The ablation switches the option off.

\paragraph{C: Model Hessian and coordinates.}
Sella starts from an approximate Hessian built from the structure alone, in which every bond, angle and dihedral is assigned a stiffness by simple rules.
AutoSella-F changes this model in three ways.

\emph{Stiffness by chemical type.}
The stiffnesses of the existing coordinates depend on more detailed chemical classes, for example the kind of single bond a rotation is about, nearly linear groups, and bonds to heavier elements and metal ions.

\emph{Non-covalent contacts.}
New terms describe interactions the original model ignores: hydrogen bonds, contacts of X--H groups with aromatic rings, and atoms close in space but far apart along the chain of bonds.
These are not new coordinates: their stiffness is computed for the atoms involved and projected onto the existing internal coordinates.

\emph{Coordinates.}
The evolution extends Sella's set of redundant internal coordinates with out-of-plane coordinates for planar groups such as carbonyls.
They describe bending of such a group out of its plane directly, which the original set captured only indirectly through dihedral angles.

The ablation restores Sella's original model and coordinates.

\paragraph{D: Hessian updates.}
A quasi-Newton optimizer refines its Hessian from pairs of a step and the resulting change of the gradient, each of which measures the curvature along that step.
AutoSella-F improves this update in three ways.

\emph{Memory.}
Sella uses only the latest pair; AutoSella-F imposes the last four at once, discarding older pairs that are inconsistent with the newest one.

\emph{Following the geometry.}
The part of the model Hessian that depends on interatomic distances, the stiffness of bonds and non-covalent contacts, is recomputed at every new geometry, and the stored pairs, measured at earlier geometries, are corrected by the same change so that they describe the curvature at the current point.

\emph{Symmetry.}
If the molecule is symmetric, the curvature along a step equals the curvature along its symmetric image, for example its mirror image.
Every stored pair is therefore also entered in its symmetric copies, which gives additional curvature information.

The ablation removes all three and returns to Sella's update from the latest pair alone.

\paragraph{E: Step control.}
Sella limits each step to a trust radius, which it enlarges or shrinks after every step according to how well the model predicted the change in energy.
AutoSella-F changes this control in three ways.

\emph{Trust radius.}
The radius starts larger, grows faster after well-predicted steps and is halved after a failed one.

\emph{Atomic displacement.}
For single molecules, a step is additionally limited so that no atom moves by more than twice the trust radius.

\emph{Step refinement.}
For single molecules, the step is refined before it is taken: the stiffness of non-covalent contacts is re-evaluated along the proposed step, and the step is recomputed with it, at most twice and without calls to the potential.

The ablation restores Sella's radius rules and removes the displacement limit and the refinement.

\subsection{AutoSella-A blocks}

\paragraph{A: Fragment handling.}
As in block B of AutoSella-F (Section~\ref{par:f-fragments}), AutoSella-A enables Sella's fragment option; in addition, the initial stiffness of the fragment position and orientation coordinates is ten times smaller ($0.005$ instead of $0.05$\,Hartree), so fragments move more freely at the start.
The ablation switches the option off and restores the original stiffness.

\paragraph{B: Model Hessian and coordinates.}
The approximate Hessian that Sella starts from is rebuilt from published stiffness rules: Badger's rule for bonds, Schlegel's for angles and torsions and Fischer's for out-of-plane bending, together with extra springs between atoms of the same molecule.
In conjugated hydrocarbons, the model also couples the stretching of neighbouring bonds, which share their $\pi$ electrons.
The set of internal coordinates gains out-of-plane coordinates averaged over atom orderings, bending coordinates for nearly linear groups, and torsions across linear chains.
The ablation removes all of these additions.

\paragraph{C: Calibration.}
The model Hessian is a sum of parts --- bonds, angles, torsions, out-of-plane terms and a few smaller ones --- each part scaled by a single factor that starts at one.
After the first six usable steps, these factors are fitted to the curvature measured along those steps, within a factor of two of their initial values, and the Hessian updates are then reapplied.
The ablation keeps all factors at their initial values.

\paragraph{D: Coordinate transformation.}
Bond lengths are replaced by a nonlinear function of the length, chosen after Badger's rule so that the energy is closer to quadratic in it.
When a step in internal coordinates is converted back to atomic positions, the conversion is updated along the way instead of being held fixed.
When a new covalent contact forms, the set of internal coordinates is rebuilt.
The ablation restores plain bond lengths, the fixed conversion and the original rebuilding rule.

\paragraph{E: Hessian update.}
The update mixes Sella's usual formula with a rank-one correction, weighted by how well each fits the latest step.
Directions in which the model has negative curvature are treated as positive, both when a step is computed and when the Hessian is updated.
The ablation restores Sella's original update.

\paragraph{F: Collision and step control.}
The initial trust radius is larger for single molecules ($0.2$) than for fragment positions and orientations ($0.1$).
Before the potential is called, a step that would push non-bonded atoms into each other is shortened, and the trust radius is adjusted for the shortened step.
The ablation restores the original radius and removes the collision check.

\paragraph{G: Gaussian process.}
From the recent energies and gradients, AutoSella-A fits a Gaussian-process model of the energy in a small space spanned by the ordinary step and recent displacements.
Minimizing this model proposes an alternative step, which is taken only if it passes checks on descent, size, predicted gain and uncertainty; otherwise the ordinary step is used.
Fitting and minimizing the model needs no additional calls to the potential.
The ablation always takes the ordinary step.

\subsection{Dependencies, results, and unsuccessful evaluations}
We supplement all single block removals with joint removals: Fable C \& E and Astra B \& C, B \& E, and B \& G.
These pairs probe the contact model and its step correction, and the physical prior's connections to calibration, curvature treatment, and the surrogate.
For all four tested pairs on validation, joint removal increases cost less than the sum of the individual removal penalties.
Thus, the single-block effects cannot be treated as independent, additive contributions.

Figure~\ref{fig:family-ablations} reports every measured variant. The full
AutoSella-F costs are 65.13\% and 64.99\% of Sella on the training and
validation splits; AutoSella-A costs are 74.51\% and 76.66\%.
The physical-model and coordinate family has the largest measured
single-removal cost increase in both series: Fable C adds 15.57 and
18.68 percentage points, while Astra B adds 8.82 points on validation;
the latter's training result fails the energy criterion. Removing Fable D
adds 9.66 and 10.16 points.

Of all evaluations, only five complete evaluations fall below the energy threshold: Fable $-A$ on training; Astra $-B$, $-B$ \& $-G$, and $-F$ on training; and Astra $-A$ on validation.
Their measured costs remain visible as red crosses.
And only three evaluations contain optimizer errors.
This indicates that described blocks represent meaningful self-contained families of improvements with little or no effect on adjacent blocks.

\section{Autoresearch with Grok 4.6}
\label{app:grok}
\begin{figure}[!htbp]
\centering
\includegraphics[width=\textwidth]{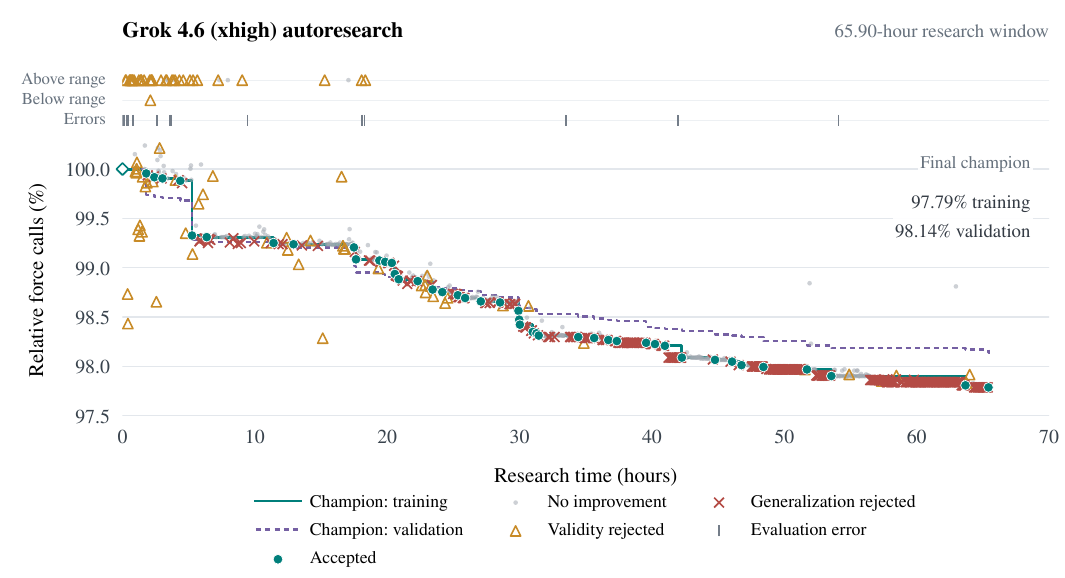}
\caption{\textbf{Autoresearch progress of Grok~4.6 (xhigh) over 66 active research hours.}
Solid and dashed lines show the current champion's training and validation force-call costs, relative to the initial Sella program.
Markers indicate candidate outcomes decided by admission gates of Algorithm~\ref{alg:ar}.
Out-of-range candidates and evaluation errors appear in the event rows above the plot.}
\label{fig:progress-grok}
\end{figure}

We additionally ran Grok~4.6 with xhigh reasoning effort through the Cursor agent harness, using the same exact setup as the main experiments.
The run recorded 900 candidates, and accepted 44 improvements.
Its final champion, AutoSella-G, was obtained at cycle~897 and reduced mean relative force-call cost to 97.79\% on training and 98.14\% on validation.
Figure~\ref{fig:progress-grok} shows the full run, which spans 66 active research hours.

On the held-out GFN2-xTB benchmarks, AutoSella-G provides little benefit on single-molecule systems and performs substantially worse on the intermolecular datasets (Table~\ref{tab:grok}).
Its mean per-molecule force-call cost is 99.27\% of Sella on Test and 100.49\% on Dipeptides, rising to 148.77\% on amino acid--ligand pairs and 239.85\% on Solvated PubChem (calculated on converged optimizations).
For Solvated PubChem, 64 optimizations do not converge within the budget and four encounter execution errors.
Given this poor GFN2-xTB performance, we did not evaluate AutoSella-G with DFT and excluded it from the main-text comparisons.
These results underscore the performance gap between the frontier models Fable and Astra and the competing Grok~4.6 model on this complex research task.

\begin{table}[!htbp]
\centering
\caption{\textbf{AutoSella-G on the four GFN2-xTB evaluation datasets.}
Relative force-call costs and recovered-energy ratios follow Eqs.~\eqref{eq:cost}--\eqref{eq:energy} and use jointly converged pairs.
Err.\ b/a counts unsuccessful Sella/AutoSella-G runs, including non-convergence within the force-call budget.}
\label{tab:grok}
\small
\setlength{\tabcolsep}{3pt}
\renewcommand{\arraystretch}{0.98}
\setlength{\aboverulesep}{1.5pt}
\setlength{\belowrulesep}{1.5pt}
\begin{tabular*}{0.94\textwidth}{@{\extracolsep{\fill}}l r l c S[table-format=3.2] S[table-format=3.2] S[table-format=1.6, group-digits=none]@{}}
\toprule
\multicolumn{4}{@{}l}{\textbf{GFN2-xTB}} & \multicolumn{2}{c}{Relative force calls (\%, Eq.~\eqref{eq:cost})\,$\downarrow$} & \\
\cmidrule(lr){5-6}
Benchmark & {$N$} & Method & {Err.\ b/a\,$\downarrow$} & {$C_{\mathrm{pooled}}$} & {$C_{\text{per-mol}}$} & {$\bar r$\,$\uparrow$} \\
\midrule
Test & 500 & AutoSella-G & 0/1 & 99.05 & 99.27 & 1.000074 \\
\cmidrule(lr){1-7}
Dipeptides & 673 & AutoSella-G & 0/0 & 99.71 & 100.49 & 1.000449 \\
\cmidrule(lr){1-7}
AA--ligand pairs & 475 & AutoSella-G & 0/2 & 140.32 & 148.77 & 1.027613 \\
\cmidrule(lr){1-7}
Solvated PubChem & 100 & AutoSella-G & 2/68 & 216.52 & 239.85 & 0.983101 \\
\bottomrule
\end{tabular*}
\end{table}

\FloatBarrier

\end{document}